\documentclass[aps,prb,twocolumn,superscriptaddress, show pacs]{revtex4-1}

\usepackage{bm, amsmath, amssymb, latexsym}
\usepackage{graphicx}
\usepackage{color}
\usepackage{dcolumn}
\usepackage{subfigure}

\begin{document}

\title{The properties of isolated chiral skyrmions  in thin magnetic films}

\author{A.~O.~Leonov}
\thanks{A.Leonov@ifw-dresden.de}
\affiliation{IFW Dresden, Postfach 270016, D-01171 Dresden, Germany}   
\affiliation{Zernike Institute for Advanced Materials, University of Groningen, \\
Groningen, 9700AB, The Netherlands}

\author{T.~L.~Monchesky}
\affiliation{Department of Physics and Atmospheric Science, Dalhousie University, Halifax, Nova Scotia, Canada B3H 3J5}

\author{N.~Romming} 
\affiliation{Department of Physics, University of Hamburg, D-20355 Hamburg, Germany}

\author{A.~Kubetzka} 
\affiliation{Department of Physics, University of Hamburg, D-20355 Hamburg, Germany}

\author{A.~N.~Bogdanov}
\thanks{ A.Bogdanov@ifw-dresden.de}
\affiliation{IFW Dresden, Postfach 270016, D-01171 Dresden, Germany}

\author{R.~Wiesendanger} 
\affiliation{Department of Physics, University of Hamburg, D-20355 Hamburg, Germany}

\date{\today}

\begin{abstract}
{ Axisymmetric solitonic states 
(\textit{chiral skyrmions}) have been predicted theoretically
more than two decades ago. However, until recently they have been 
observed in a form of skyrmionic condensates (hexagonal lattices 
and other mesophases).
In this paper we report experimental and theoretical investigations
of \textit{isolated} chiral skyrmions  discovered in
PdFe/Ir(111) bilayers two years ago 
(Science \textbf{341} , 636 (2013)).
The results of spin-polarized scanning tunneling microscopy  
analyzed within the continuum and discrete models provide 
a consistent description of isolated skyrmions in thin layers.  
The existence region of chiral skyrmions is restricted 
by strip-out instabilities at low fields and a collapse 
at high fields.
We demonstrate that the same equations describe axisymmetric
localized states in all condensed matter systems with 
broken mirror symmetry, and thus our findings establish 
basic properties of isolated skyrmions common for chiral 
liquid crystals, different classes of noncentrosymmetric 
magnets, ferroelectrics, and multiferroics.
}
\end{abstract}

         
\maketitle

\vspace{3mm}

\section{Introduction}

\begin{figure}
\includegraphics[width=1.0\columnwidth]{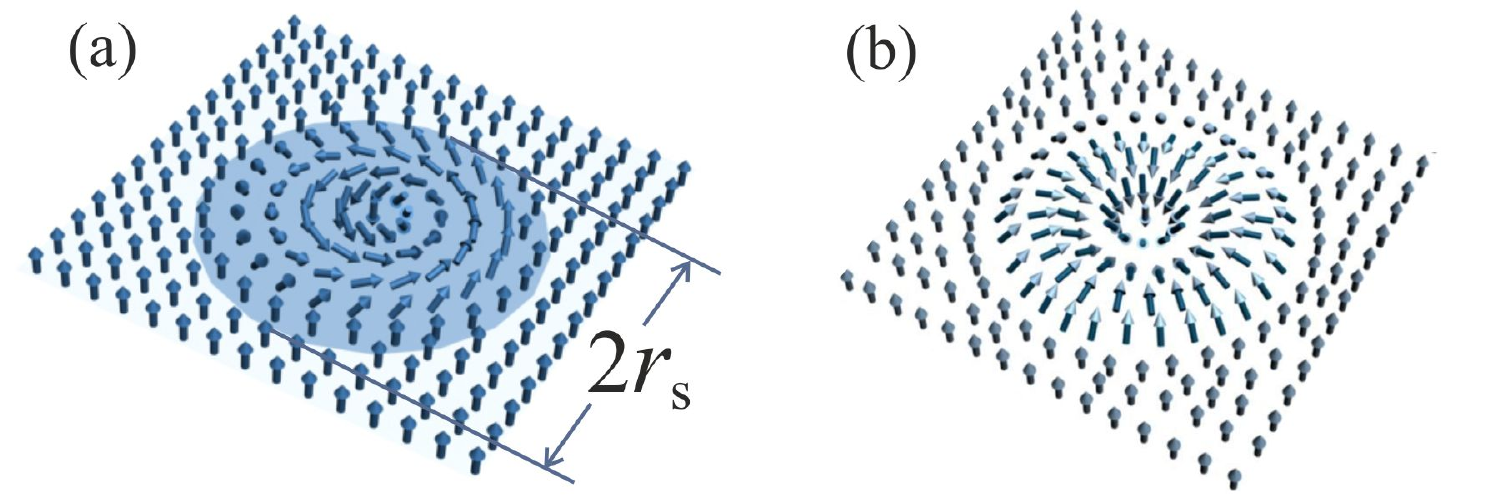}
\caption{ 
(color online). Axisymmetric isolated skyrmions:
(a) in cubic helimagnets and uniaxial ferromagnets
with $D_n$ symmetry; (b)
in uniaxial ferromagnets with  $C_{nv}$  symmetry \cite{JETP89}.
\label{structure1}
}
\end{figure}

\begin{figure}
\includegraphics[width=1.0\columnwidth]{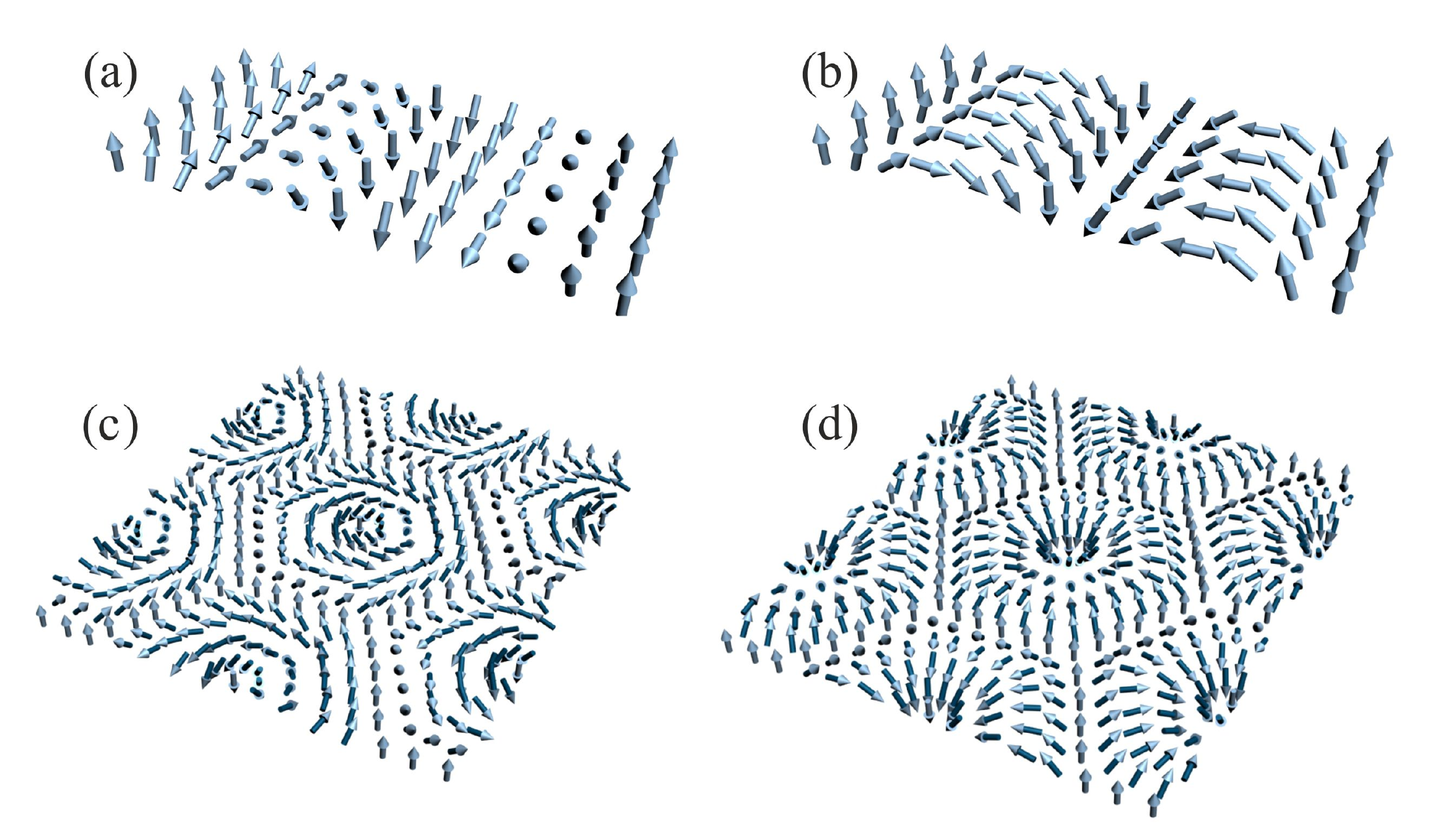}
\caption{ 
(color online). Basic modulated phases in chiral ferromagnets:
one-dimensional \textit{helicoids} (a)
and \textit{cycloids}  (b) and two-dimensional
\textit{skyrmion lattices} (c,d). 
Bloch-type modulations (a,c) arise 
in cubic helimagnets and ferromagnets with $D_n$ symmetries;
N\'eel-type modulations (b,d) are attributed to
 uniaxial ferromagnets with $C_{nv}$  symmetries \cite{JETP89}.
\label{structure1a}
}
\end{figure}

Long-period homochiral magnetization modulations 
(\textit{helical} phases) \cite{Dz64} and axisymmetric 
solitonic patterns (\textit{vortices} or \textit{skyrmions}) 
\cite{JETP89,JMMM94,JPCS11} are two types of unconventional  magnetic states 
attributed solely to magnetic  compounds with broken inversion 
symmetry and distinguish them from common (achiral) magnetic 
materials (Figs. \ref{structure1}, \ref{structure1a}).
Both, extended chiral modulated phases and localized skyrmionic
states are stabilized by specific \textit{Dzyaloshinskii-Moriya} (DM)
interactions arising in chiral magnets owing to
their crystallographic handedness \cite{Dz64}.
In the micromagnetic energy functionals of noncentrosymmetric 
ferromagnets these interactions are described by energy contributions 
linear in the first spatial derivatives of the magnetization 
$\mathbf{M}$ (\textit{Lifshitz} invariants) \cite{Dz64}
\begin{eqnarray}
 M_i \frac{\partial M_j}{\partial x_k} 
- M_j \frac{\partial M_i}{\partial x_k} .
\label{lifshitz}
\end{eqnarray}

Axisymmetric localized structures (Fig. \ref{structure1})
are related to multidimensional topological solitons 
with nonsingular internal structure and finite energy 
\cite{Rajaraman87}. 
These particle-like objects  are of special interest 
in fundamental physics and mathematics 
\cite{Manton04, Skyrmion10, Melcher15}.
In most nonlinear physical systems, multidimensional solitons 
(skyrmions) can exist only as dynamic excitations while 
static solutions are unstable and collapse spontaneously 
into topological singularities \cite{Derrick64}.

In nonlinear field theory, the existence and stability of
skyrmion solutions is provided by special terms in the
energy functionals. 
More than five decades ago T. H. Skyrme 
introduced into the nonlinear field model
an interaction term with higher order 
spatial derivatives that stabilize
two- and three-dimensional 
topological nonsingular solitons (now commonly addressed as 
\textit{skyrmions}) \cite{Skyrme62}.
Since that time, field theorists have been intensively 
investigating this family of solitons (skyrmions) within  
the Faddeev-Skyrme and kindred models \cite{Skyrme62, Manton04,Leonov15a}.

Lifshitz invariants of type (\ref{lifshitz}) provide the only 
known alternative to the Skyrme mechanism
that yield regular solutions for axisymmetric skyrmions
\cite{JMMM94,JETPL95,Nature06}.
These invariants arising in noncentrosymmetric condensed matter systems 
(including chiral magnets, liquid crystals, multiferroics, and
nanolayers of magnetic metals with interface induced Dzyaloshinskii-Moriya
interactions) introduce a unique class of materials where mesoscopic skyrmions 
can be induced and manipulated.  

In a broad range of applied magnetic fields and temperatures 
isolated skyrmions condense into hexagonal lattices 
\cite{JMMM94,Wilhelm11,Wilson14} or other types of two-dimensional
modulated states \cite{Wilhelm12,Keesman15}.
During the last years, intensive experimental efforts 
have been  undertaken to find  indications of hexagonal skyrmion
lattices in different groups of chiral ferromagnets (see e.g 
\cite{Lamago06,Pappas09,Muhlbauer09,Yu10,Heinze11,Yu12,Wilson12,
Huang12,Kezsmarki15,Tokunaga15} and bibliography in \cite{Wilson14}).
Particularly, direct observations of skyrmion lattices have been reported
in free standing nanolayers of cubic helimagnets in \cite{Yu10} 
(and the following papers of this group \cite{Yu12,Seki12}). 
These results reveal axial symmetry and homochirality 
of the embedded skyrmions, and observed properties of skyrmion lattices
were found to be in close correspondence with theoretical results.
To date the LTEM studies of confined cubic helimagnets 
have focused  on  the skyrmion condensates
(skyrmion lattices and clusters) \cite{Yu10,Yu12,Seki12,Tokunaga15}.
Spin-polarized scanning tunneling microscopy (SP-STM)
has been able to identify isolated skyrmions in the saturated states 
of PdFe/Ir(111) films \cite{Romming13}, and subsequently 
resolve their internal structure  \cite{Romming15}.

In this paper we present detailed experimental and theoretical
investigations of axisymmetric isolated skyrmions in thin magnetic
films.

In the theoretical part we develop a consistent theory of chiral
skyrmions in thin magnetic layers (Sec. IIA). In Sec. IIB we
apply the qualitative theory of differential equations 
to expound main features of isolated chiral
skyrmions and elucidate their physical nature, investigate
the conditions of the elliptical instability at low fields
and calculate within the discrete model the skyrmion 
collapse field. In Sec. IIC we construct the phase diagram 
of the solutions for isolated skyrmions.

In the experimental part we present the detailed evolution
of isolated skyrmions in PdFe/Ir(111) bilayers from the
strip-out at low fields to the collapse at high fields.

\section{Theory}

A phenomenological theory of chiral modulations
in noncentrosymmetric magnetic crystals has been developed
by I. Dzyaloshinskii in 1964 \cite{Dz64}. These papers
also include analytical solutions for one-dimensional 
chiral modulations (\textit{helicoids} and \textit{cycloids}).
Theory of isolated skymions and skyrmion lattices  
in bulk noncentrosymmetric ferromagnets has been developed 
in \cite{JMMM94,pss94}.
Theoretical investigations of chiral modulations in bulk and confined
noncentrosymmetric ferro- and antiferromagnets have been carried out
in many of the papers discussed in Ref. \cite{Wilson14}.

\subsection{The micromagnetics of chiral modulations}

\subsubsection{Energy functional and symmetry}

In this paper we investigate isolated skyrmions in a thin layer
of a noncentrosymmetric ferromagnet.
As a model we consider a thin plate infinite 
along the $x-$ and $y-$ axes and
of thickness $L$ along the $z-$ axis.
In the following sections we specify the model and
discuss its limitations.
For a  film of a noncentrosymmetric uniaxial ferromagnet in the applied
magnetic field $\mathbf{H}^{(e)}$ perpendicular to the film surface,
the micromagnetic energy density written within terms quadratic in the
components of the magnetization vector $\mathbf{M}$
has the following standard form \cite{Dz64}:
\begin{eqnarray}
w &=& A (\mathbf{grad} \, \mathbf{m})^2 +
w_D(\mathbf{m}) - K  (\mathbf{m} \cdot \mathbf{n} )^2 
\nonumber \\ 
&-& \mu_0 M H^{(e)} \mathbf{m} \cdot \mathbf{n}
- \frac{1}{2} \mu_0 M \mathbf{m} \cdot \mathbf{H}^{(d)},
\label{energy}
\end{eqnarray}
where $A$ is the exchange stiffness constant, 
$K$ is the uniaxial anisotropy constant,
$\mathbf{H}^{(d)}$ is the demagnetizing field,
\begin{eqnarray}
\mathbf{m}= \mathbf{M}/|\mathbf{M}| 
= (\sin \theta \cos \psi, \sin \theta \sin \psi, \cos \theta)
\label{mvector}
\end{eqnarray}
is the reduced magnetization,
$\mathbf{n}$ is the unity vector directed perpendicular to 
the film surface.

The Dzyaloshinskii-Moriya energy density
$w_D$ is composed of Lifshitz invariants (\ref{lifshitz}):

\begin{eqnarray}
\mathcal{L}_{ij}^{(k)} = m_i \frac{\partial m_j}{\partial x_k} 
- m_j \frac{\partial m_i}{\partial x_k} .
\label{lifshitz1}
\end{eqnarray}

The functional forms of  energy density $w_D$ are determined by
crystallographic symmetry of a noncentrosymmetric magnetic crystal
and are listed in Eqs. (\ref{lifshitz2}), (\ref{lifshitz4}).
Lifshitz invariants (\ref{lifshitz1}) favour spatial modulations
with a fixed rotation sense along the $x_k$ directions  \cite{Dz64}.
A competition between  the chiral energy $w_D$ and other
energy contributions leads to the formation of isolated chiral states
\cite{JETP89,JMMM94} and spatially modulated magnetic phases \cite{Dz64,JMMM94}.

The Euler equations for energy functional (\ref{energy}) together
with Maxwell's equations,
\begin{eqnarray}
\mathrm{rot}  \mathbf{H}^{(d)} = 0, \quad
\mathrm{div} \left[ \mathbf{H}^{(d)} +\mu_0 \mathbf{M} \right] =0,
\label{Maxwell}
\end{eqnarray}
yield solutions for different types of chiral modulations
(Figs. \ref{structure1}, \ref{structure1a}, \ref{fig:structure2}).

\subsubsection{ Demagnetization effects}

Generally the equilibrium modulated patterns $\mathbf{m} (\mathbf{r})$  
in a chiral magnet are derived by numerically solving 
the above set of nonlinear differential equations including 
non-local stray field calculations \cite{JMMM94,Kiselev11}.
Contrary to soft magnetic materials where demagnetization fields sufficiently influence
the equilibrium magnetic states \cite{Hubert98}, in chiral magnetic materials
the DM interactions strongly suppress these effects \cite{Kiselev11}. As a result
in many practical cases  a magnetostatic problem is reduced to analytical solutions
\cite{Tu71,JMMM94,Kiselev11}, and the stray-field energy 
can be expressed as local energy contributions 
in energy functional (\ref{energy}) \cite{JMMM94,Kiselev11}.

It was also found that for one-dimensional modulations
and two-dimensional axisymmetric structures, the internal
stray-field energy has a local character \cite{Hubert98,JMMM94}.
Particularly, for ferromagnets with C$_{nv}$ symmetry 
the internal stray-field energy can be taken into account 
by the following redefinition of the anisotropy constant,
\begin{eqnarray}
K \rightarrow  K + K_d, \quad K_d = \mu_0 M^2/2.
\label{Keff}
\end{eqnarray}

\subsubsection{ The equations for axisymmetic skyrmions}

We introduce cylindrical coordinates for the spatial
variable $ \mathbf{r} =  ( r \cos \varphi, r \sin \varphi, z)$
and consider magnetic patterns homogeneous along the 
$z$-axis with the magnetization antiparallel to the applied
field in the center ($\theta  = \pi$ for $r = 0$)
and approaching  the parallel orientation  when the distance
from the center approaches infinity 
($\theta \rightarrow 0$ for $r \rightarrow \infty$).
For $\theta (\rho, \varphi)$, $\psi (\rho, \varphi)$
the energy functional (\ref{energy}) is reduced to the 
following form:
\begin{eqnarray}
w  &=& A \left[ \theta_{r}^2
+\frac{1}{r^2} \theta_{\varphi}^2
+ \sin^2 \theta \left( \psi_{r}^2
+\frac{1}{r^2} \psi_{\varphi}^2 \right) \right]
+ w_D 
\nonumber \\ 
&-&  K \cos^2 \theta - \mu_0 M H^{(e)} \cos \theta
-  \mu_0 M \mathbf{m} \cdot \mathbf{H}^{(d)},
\label{energy2}
\end{eqnarray}
and the Dzyaloshinskii-Moriya energy functionals 
$w_D (\theta, \psi, r, \varphi)$ 
are listed in Eqs. (\ref{lifshitz2}), (\ref{lifshitz4}).

The equations minimizing energy (\ref{energy2}) include
rotationally symmetric solutions, 
\begin{eqnarray}
\theta =\theta (\rho), \quad \psi = \psi (\varphi), \quad
\mathbf{H}^{(d)} = \mathbf{H}^{(d)} (\rho).
\label{axial}
\end{eqnarray}
Analytical solutions $\psi =\psi (\varphi)$
for uniaxial noncentrosymmetric ferromagnets \cite{JETP89}  
and cubic helimagnets (Figs. \ref{structure1}, \ref{structure1a})
are listed in Eq. (\ref{lifshitz3}). 

To date, only two types of skyrmionic states from this list have been identified
in chiral ferromagnets by direct experimental observations:
skyrmionic patterns with Bloch-type modulations (Fig. \ref{structure1} a)
\begin{eqnarray}
\mathbf{m} = \vec{e}_{\varphi} \sin \theta(\rho) + \vec{e}_{z} \cos \theta (\rho)
\label{blochskyrmion}
\end{eqnarray}
have been observed in free standing nanolayers
 of cubic helimagnets (see e.g. \cite{Yu10,Yu12,Seki12}),
and skyrmion lattices with N$\acute{e}$el-type modulations (Fig. \ref{structure1} b)
\begin{eqnarray}
\mathbf{m} = \vec{e}_{\rho} \sin \theta(\rho) + \vec{e}_{z} \cos \theta (\rho)
\label{neelskyrmion}
\end{eqnarray}
have been observed in Fe/Ir(111) and PdFe/Ir(111) nanolayers
\cite{Heinze11,Romming13,Romming15,Bergmann15} and in the 
rhombohedral ferromagnet GaV$_4$O$_8$ with  C$_{3v}$ symmetry 
\cite{Kezsmarki15}.

The first direct observations of isolated skyrmions 
 have been reported in PdFe/Ir(111) nanolayers \cite{Romming13}. 
These chiral solitonic structures have been investigated in a broad range 
of applied fields \cite{Romming13,Romming15}.

After integration with respect to $\varphi$, the total energy  
$\mathcal{F}$ for an isolated skyrmion of Bloch- and N$\acute{e}$el-type 
in an applied magnetic field perpendicular to the film surface can 
be reduced to the following form:
\begin{eqnarray}
\textstyle
\mathcal{F} = 2 \pi\int_0^{\infty} f (\theta, r) r d r.
\label{totalenergy}
\end{eqnarray}
Here $f (\theta, r) = w (\theta, r)- w (0)$ is the difference between
the skyrmion energy density and that of the saturated state,
$w(0) = - K - \mu_0 M H$ :
\begin {eqnarray}
f (\theta, r) &=& A \left( \theta_{r}^{2} + \frac{1}{r^2} \sin^2 \theta \right)
- D \left(\theta_{r} +\frac{1}{r} \sin \theta \cos \theta \right)
\nonumber \\
&+& K \sin^2 \theta + \mu_0 H \left(1- \cos \theta \right).
\label{energy3}
\end{eqnarray}
In Eq. (\ref{energy3}) $H \equiv H_z$ is the perpendicular component 
of the internal magnetic field that differs from the applied external field 
($H^{(e)}$) due demagnetization field of the film
surface \cite{Hubert98}. For rather thick  films ($ d \geq r_s$) 
$H = H^{(e)} - \mu_0 M$ and for ultrathin films
$H \approx   H^{(e)}$.

The Euler equation for energy functional (\ref{energy3}), 
\begin {eqnarray}
&A& \left(\theta_{r r} + \frac{1}{r} \theta_{r} -\frac{1}{r^2} \sin \theta \cos \theta \right)
+\frac{D}{r} \sin^2 \theta \quad
\nonumber \\
&-& K \sin \theta \cos \theta - \mu_0 M H \sin \theta = 0,
\label{equation0}
\end{eqnarray}
with boundary conditions
\begin{eqnarray}
\theta (0) = \pi, \quad \theta (\infty) = 0, 
\label{boundary}
\end{eqnarray}
yields the equilibrium structure of isolated axisymmetric skyrmions
\cite{JETP89,JMMM94}.
Note that for  N$\acute{e}$el-type  skyrmions $K$ includes
the stray energy contribution  (\ref{Keff}).

\begin{figure*}
\includegraphics[width=2.0\columnwidth]{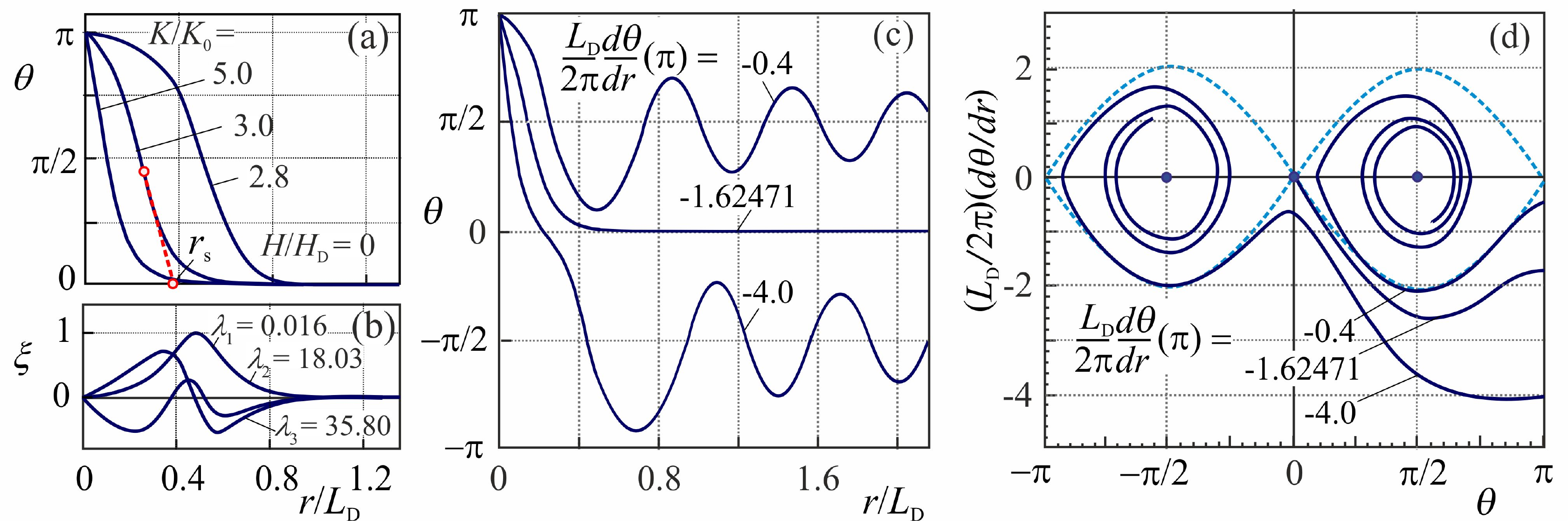}
\caption{ 
(color online). 
(a) Typical  localized solutions of the boundary value
problem with magnetization profile $\theta (\rho)$.
 The first three excitation modes $\zeta_i (\rho)$
with positive eigenvalues $\lambda_i$ (b) for 
the solution for $K/K_0$ = 2.8.
``Shooting trajectories $\theta (\rho)$ of the Cauchy
problem (c) and corresponding phase trajectories 
$\theta_{\rho} (\theta)$ (d).
\label{solutions}
}
\end{figure*}

Dimensionless variables
\begin{eqnarray}
\rho = 2 \pi r/L_D, \quad  h = H/H_D, \quad k = K/K_0,
\label{units5}
\end{eqnarray}
are commonly used in recent papers to describe modulated states
in uniaxial chiral ferromagnets and cubic helimagnets
(see e.g. \cite{Butenko10,Huang12,Wilson14,Romming15}).
Here we use the characteristic parameters of a uniaxial chiral ferromagnet 
\cite{JMMM94,Butenko10}:
\begin{eqnarray}
 L_D = \frac{4 \pi A}{|D|}, \quad \mu_0 H_D =\frac{ D^2}{2A M},
\quad K_0 = \frac{D^2}{4A} . \qquad 
\label{unitsLD}
\end{eqnarray}
$L_D$ is the period of a helix at zero field and zero anisotropy, 
$H_D$ is the saturated field and  $K_0$  is  the critical anisotropy 
(\ref{cone1}).

With variables (\ref{units5}), the equation for axisymmetric 
skyrmions (\ref{equation0}) is reduced to the following form:
\begin {eqnarray}
\theta_{\rho \rho} + \frac{\theta_{\rho}}{\rho} 
&-&\frac{1}{\rho^2} \sin \theta \cos \theta
+ \frac{2\sin^2 \theta}{\rho} \nonumber
\\ 
&-& k \sin \theta \cos \theta - h \sin \theta =0,
\label{equation1}
\end{eqnarray}
with boundary conditions (\ref{boundary}).

\subsection{ Solutions for axisymmetric skyrmions}

\begin{figure}
\includegraphics[width=1.0\columnwidth]{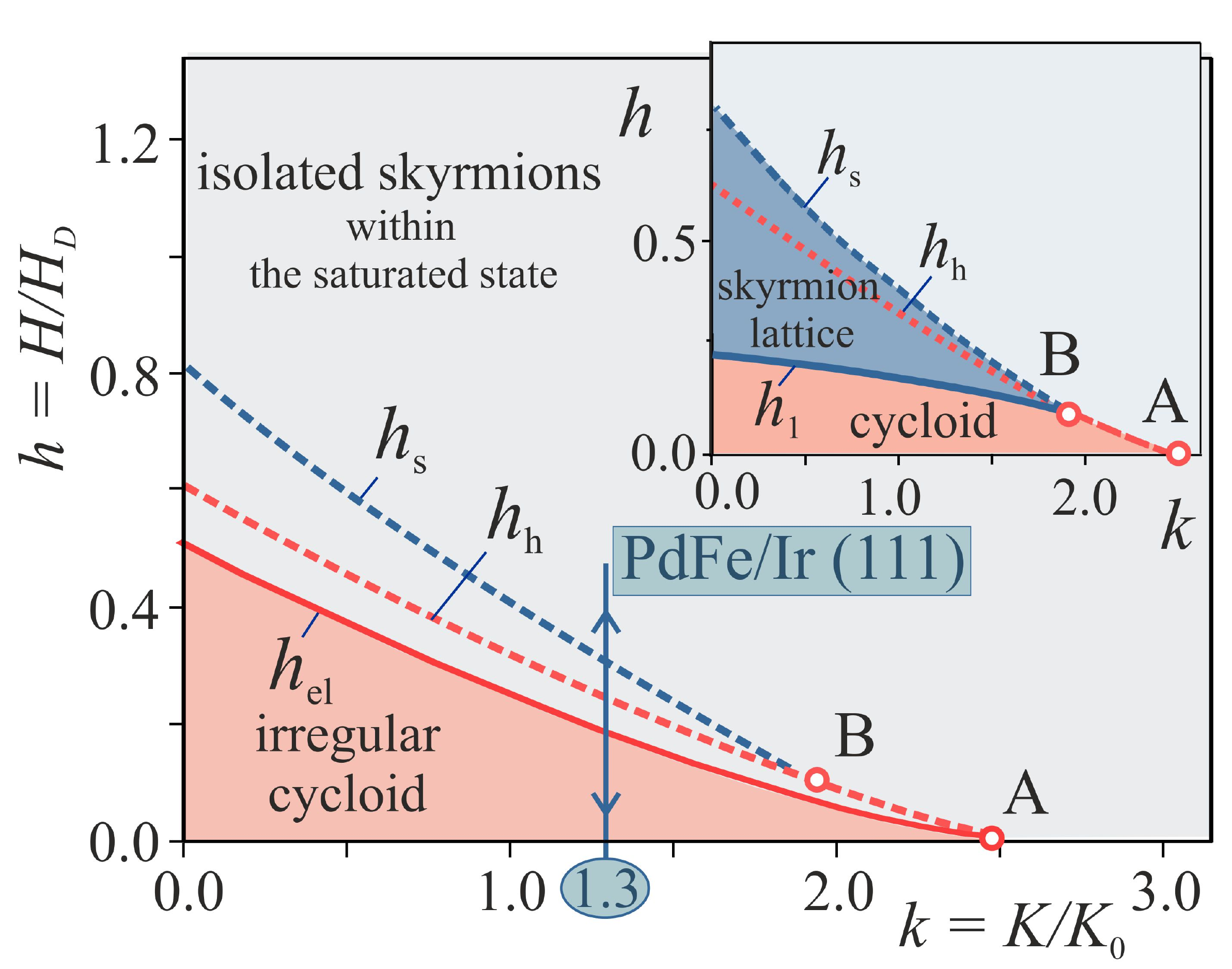}
\caption{
(color online).
In the phase diagram in variables $k$ and $h$
the existence area of metastable isolated skyrmions is restricted
by the strip-out critical line $h_{el} (k)$.
The inset shows the regions of global stability
of the modulated (helicoidal and skyrmion lattice) and
the spatially homogeneous saturated phases
(for details see \cite{Wilson14}).
\label{phasediagrams}
}
\end{figure}

The equilibrium skyrmion profiles $\theta (\rho)$
are derived by solving the boundary value problem 
(\ref{equation0}) and (\ref{boundary}) with a finite-difference 
method \cite{JMMM94}.
Typical solutions of Eq. (\ref{equation0}) are plotted 
in Fig. \ref{solutions},
and the existence areas for isolated skyrmions
 are indicated in the phase diagram of the solutions
(Fig. \ref{phasediagrams}).

The solutions $\theta(\rho)$ are linear near the skyrmion axis
($(\pi - \theta ) \propto  \rho $ for $\rho \ll 1 $) and decay exponentially
at high distances from the center ($ \rho \gg 1 $)
$\theta  \:\propto \ \exp{\left(- \rho \sqrt{k+h}  \right)}/\sqrt{\rho}$.
%
%

Usually the functions $\theta (\rho)$ have arrow-like shape with the steepest
slope at the  center of the skyrmion ($r = 0$). 
They transform into bell-shape profiles only near the critical line $H_{el}$ . 
In micromagnetism, the  characteristic size of a localized magnetization profile 
$\theta (\rho)$ is defined as \cite{Hubert98}
\begin{eqnarray}
r_s = r_0 - \theta_0  \left( d\theta /d r \right)^{-1}_{ r = r_0}, 
\label{size}
\end{eqnarray}
where $(r_0, \theta_0$) is the inflection point of  the profile $\theta (r)$
(Fig. \ref{solutions} a).

The basic properties of the solutions for Eq. (\ref{equation0})
have been investigated in \cite{JMMM94,pss94}. 
Theories of static chiral skyrmions in different classes of bulk
and confined chiral magnets have been developed in a number of
studies (e.g., \cite{Butenko10,Lin13,Rohart13,Kim14,Wilson14}).
Numerical solutions for isolated skyrmions 
in nanodots and other confined chiral magnets have been derived 
in a large number of recent works 
(e.g., \cite{Butenko09,Leonov14a,Sampaio13,Keesman15}).
Also, dynamical properties including current-induced movement of skyrmions 
have been intensively investigated by numerical simulations of
the Landau-Lifshitz-Gilbert equation (e.g., Refs. \cite{Sampaio13,Iwasaki13,Zhang15} 
and the bibliography in a review paper \cite{Nagaosa13}). 
The results of these numerical simulations demonstrate a rich spectrum 
of magnetic states characteristic for chiral skyrmions and various 
scenarios of their evolution under the influence of 
applied fields \cite{Sampaio13,Iwasaki13,Zhang15}.
Particularly, in confined uniaxial helimagnets the applied field induces modulated 
textures with different number of skyrmions, elongated, and half skyrmions \cite{Keesman15}.

The results of numerical simulations for stationary and moving skyrmions, however, 
still require substantial analytical analysis and physical
comprehension.
The qualitative theory of nonlinear differential equations together 
with other analytical methods provide effective tools to gain important 
insight into the physics of chiral skyrmions and establish 
mathematical relations between them and other types of magnetic solitons.

\subsubsection{Visualization of solutions on  the ($\theta, \theta_{r}$) phase plane}

Solutions $\theta (r)$  of the boundary value problem  (\ref{boundary})
can be derived by solving the auxiliary Cauchy initial value problem 
for equation (\ref{equation0}),
\begin {eqnarray}
 \theta (0) = \pi,  \quad \theta_r (0) = -a.
\label{cauchy}
\end{eqnarray}
For illustration we consider the Cauchy problem given by
(\ref{equation0}) and  (\ref{cauchy}) for $H = 0$ 
and $\varkappa =  \pi D/(4 \sqrt{A K}) = 0.8$ (\ref{kappa}).
The calculated  profiles $\theta (r, a)$ and the corresponding curves
 $\theta_r (\theta)$ in the interval [$ 0.4< a < 4.0$] are plotted in
 Fig. \ref{solutions} (c),(d).
Most of curves $\theta(r,a)$ oscillate near lines $\theta_{1,2} = \pm \pi/2$,
the maximum values of $w_a = K\sin^2 \theta$, and the corresponding profiles
 $\theta_r (\theta)$
spiral around the \textit{attractors}, points ($\pm \pi/2, 0$).  
Among these curves there is a singular line (with $a = 1.62471$) which ends in the 
\textit{saddle} point ($0, 0$) and, thus, represents a solution of the
boundary value problem for isolated skyrmions.

%

The visual representation of the solutions for the auxiliary Cauchy
problem (\ref{equation0}), (\ref{cauchy}) as 
parametrized profiles $\theta(r, a)$ (Fig. \ref{solutions} (c)) 
and $\theta_r (\theta)$ curves in ($\theta, \theta_r$) phase plane 
(Fig. \ref{solutions} (d)) reveal mathematical regularities 
in the formation of the localized  states.

To demonstrate a crucial role of the DM interactions in the stabilization
of chiral skyrmions, in the following we compare the phase portrait 
in Fig. \ref{solutions} (d) with special cases of model (\ref{energy3})
with $D$ = 0.

\begin{figure}
\includegraphics[width=1.0\columnwidth]{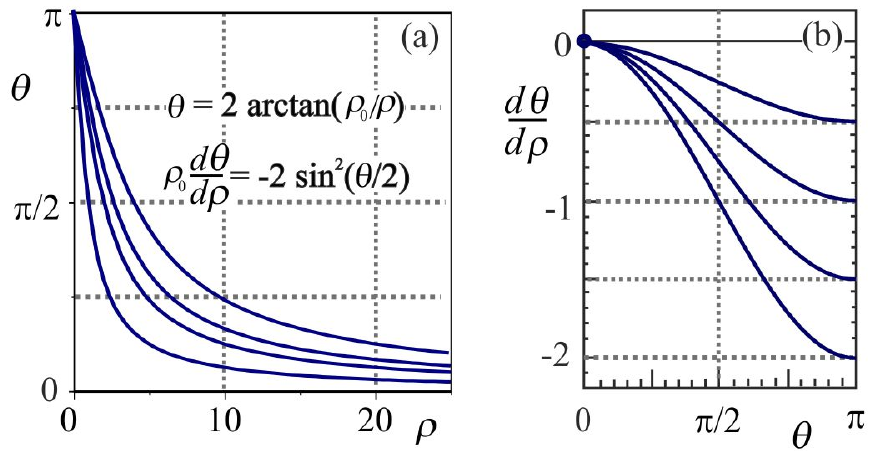}
\caption{
(color online).
Magnetization profiles for Belavin-Polyakov instantons  (a)
and the corresponding phase portraits of the solutions
(b).
\label{fig:belavin}
}
\end{figure}

\textit{Isotropic ferromagnets} ($D = K = H =0$).
The Euler equations for energy functional of an isotropic ferromagnet 
$ w = A (\mathbf{grad} \, \mathbf{m})^2$
yield rigorous analytical solutions for axisymmetric skyrmions 
$\theta(r)$, $\psi(\varphi)$ derived by Belavin and Polyakov \cite{Belavin75}
\begin {eqnarray}
 \psi = N \varphi +\alpha, \quad  \tan \left(\theta/2 \right) = ( \delta/r)^N, 
\label{instanton1}
\end{eqnarray}
where $\alpha$ and $\delta > 0$ are arbitrary values and $N$ are positive integers.
The  energy (\ref{totalenergy}) for solutions (\ref{instanton1}) 
$\mathcal{F}_0 = 8 \pi A N $, does not depend
on values $\delta$ and $\alpha$ \cite{Belavin75}.
For $N = 1$ a set of magnetization profiles $\theta (r/\delta)$ (\ref{instanton1})
and phase portrait trajectories $\theta_{r} (\theta)$ 
\begin {eqnarray}
 \theta = 2 \arctan ( \delta/r), \quad   \delta \theta_{r} = - 2 \sin^2 (\theta/2),
\label{instanton}
\end{eqnarray}
are plotted in Fig. \ref{fig:belavin}. For $\delta > 0$, the
curves $\theta_r (\theta)$ start in points ($\pi, -2/\delta$)
and end in the saddle point ($0, 0$). However, any anisotropy or magnetic field
will destabilize this solution.

\textit{Uniaxial centrosymmetric ferromagnets} ($D = 0$). In this case
Eq. (\ref{equation0}) has no stable solutions for isolated skyrmions.
For $H > 0$ all phase trajectories $\theta_r (\theta)$ spiral around
attractor ($-\pi/2, 0$). For $H< 0$   Eq. (\ref{equation0}) has radially
unstable solutions for isolated skyrmions as proved by Derrick-Hobart 
theorem (For details see \cite{JMMM94,Derrick64}).

\subsubsection{Derrick scaling identities and a virial theorem for chiral skyrmions}

Analysis of skyrmion energy $\mathcal{F}$ (\ref{totalenergy}) 
under scaling  transformations offers further important insight into
the physics of chiral skyrmions.
We consider a family of functions  $\vartheta (r) = \vartheta (r/\eta)$
obeying the boundary conditions (\ref{boundary}). 
Here $\eta > 0$ is an arbitrary constant describing uniform compressions  
($0 < \eta < 1$) or expansions ($\eta >1$) of profile $\vartheta (r)$.
For rescaled functions  $\vartheta (r) = \vartheta (r/\eta)$, 
the skyrmion energy $\widetilde{\mathcal{F}}$  (\ref{totalenergy})
can be expressed as a function of $\eta$:

\begin {eqnarray}
 \widetilde{\mathcal{F}}(\eta) = \mathcal{E}_e - \mathcal{E}_D \eta 
+ \mathcal{E}_0 \eta^2.  \quad 
\label{energyscale}
\end{eqnarray}
The values of the exchange ($\mathcal{E}_e $), 
Dzyaloshinskii-Moriya ($\mathcal{E}_D $), and
potential ($\mathcal{E}_0 $) energy contributions  
for profile $\vartheta (r)$ (\ref{totalenergy})
are given as follows:
\begin {eqnarray}
\textstyle
 \mathcal{E}_e = 2\pi A \int_0^{\infty}
\left( \vartheta_{\xi}^{2} 
+ \frac{1}{\xi^2} \sin^2 \vartheta \right) \xi d \xi  \equiv A \alpha_1,
\nonumber
\label{energyE}
\end{eqnarray}
\begin {eqnarray}
\textstyle
 \mathcal{E}_D = 2 \pi |D| \int_0^{\infty}
\left( \vartheta_{\xi}
+ \frac{1}{\xi} \sin \vartheta \cos \vartheta \right) \xi d \xi  \equiv |D| \alpha_2,
\label{energyD}
\nonumber
\end{eqnarray}
\begin {eqnarray}
\textstyle
 \mathcal{E}_0 = 2 \pi \int_0^{\infty}
\left[ K \sin^2 \vartheta 
+\mu_0 M H \left(1 -\cos \vartheta \right) \right]\xi d \xi
\label{energy0}
\end{eqnarray}
or $\mathcal{E}_0 = K  \alpha_3 + \mu_0 M H \alpha_4$, where
$\alpha_i$ are the numerical coefficients given by 
 the values of the integrals in Eqs. (\ref{energy0}). 

Eq. (\ref{energyscale}) shows that the DM
energy plays a crucial role in stabilizing skyrmions
\cite{JETP89,JETPL95}. In centrosymmetric ferromagnets 
($\mathcal{E}_D = 0$) isolated skyrmions are unstable with respect
to compression and collapse into a singular line ($\eta \rightarrow 0$)
(\textit{Derrick-Hobard} theorem \cite{Derrick64}). 
Skyrmion solutions that minimize the free energy (\ref{energyscale})
only occur for nonzero Dzyaloshinskii-Moriya energy contributions.

\textit{Ansatz solutions}.
Potential $\widetilde{\mathcal{F}}(\eta)$ (\ref{energyscale}) 
has a convenient form for analysis of skyrmion solutions with 
trial functions of type $\vartheta = \vartheta (\rho/\eta)$ 
that obey the boundary conditions (\ref{boundary}).
Particularly, a linear ansatz 
\begin {eqnarray}
 \vartheta =  \pi [1 - (r/\eta)] \quad (r < \eta), 
\quad 
\vartheta = 0 \quad  (r > \eta), \qquad
\label{linear}
\end{eqnarray}
has been used in Ref. \cite{JETP89} to introduce the phenomenon of
chiral skyrmions. The ansatz,
\begin {eqnarray}
 \vartheta (r/\eta) = 4 \arctan \left[ \exp{ \left(-r/\eta \right) } \right],
\label{ansatzwall}
\end{eqnarray}
based on solutions for isolated 360$^\circ$ Bloch walls \cite{Hubert98} 
provides a good fit  to the solutions of Eq. (\ref{equation0}).
In Ref. \cite{Romming15}, magnetization profiles for isolated skyrmions
have been fitted by a combination of functions of type (\ref{ansatzwall}).
For the trial function $\vartheta (r/\eta)$ in Eq. (\ref{ansatzwall}),
the total energy (\ref{energyscale}) can be written as

  $\mathcal{F} (\eta) /(2\pi) = 4.31 A + (1.59 K + 1.39 \mu_0 M H )\eta^2 -3.02 D \eta $.
For zero anisotropy ($k = 0$)  this ansatz yields 
the transition field into the skyrmion lattice
$h_s = H_s/H_D = 0.760$ (cf. with the rigorous value $h_s = 0.801$
and $h_s = 0.675$ for the linear ansatz (\ref{linear}) \cite{JETP89}).

For $\widetilde{\mathcal{F}}(\eta)$ (\ref{energyscale}) the equilibrium skyrmion size is
\begin {eqnarray}
\eta_0 = \frac{L_D}{2 \pi } \: \frac{\alpha_2}{ \alpha_3 k + 2 \alpha_4 h },
\label{energyeta0}
\end{eqnarray}
expressed as a ratio of the Dzyaloshinskii-Moriya to the potential energy contributions
for the trial function.

The \textit{virial} theorem for isolated axisymmetric skyrmions is 
derived by integration of the Euler equation (\ref{equation0}). 
Partial integration leads to the following \textit{virial} relation 
 between the equilibrium values of the potential 
and DM energies \cite{pss94}
$\mathcal{\bar{E}}_0 = 2 \mathcal{\bar{E}}_D$
where $\mathcal{\bar{E}}_0$, $\mathcal{\bar{E}}_D$ 
are  the integrals (\ref{energy0})
calculated for the solutions of 
Eq. (\ref{equation0}), $\vartheta = \theta(\rho)$.

\subsubsection{Radial stability and collapse at high field (discrete model)}

The stability of the solutions $\theta (\rho)$  
of the boundary value problem (\ref{equation0}), 
(\ref{boundary}) under small radial distortions $\xi (\rho)$
($\xi (0) = \xi (\infty)  = 0$) 
has been investigated in \cite{JMMM94}.
This problem is reduced to the spectral problem for 
the perturbation energy functional \cite{JMMM94}.
By numerically solving the eigenvalue problem for
this functional, the radial stability of isolated
skyrmions has been established in a broad range
of the control parameters $(k, h)$ \cite{JMMM94}.
Contrary to magnetic bubbles, which collapse with  finite radii 
at certain critical fields \cite{Hubert98}, the solutions of
the boundary value problem (\ref{equation0}) and (\ref{boundary}) 
(Fig. \ref{solutions}) exist at arbitrary high fields. 
In increasing fields their sizes gradually decrease and asymptotically approach  zero.

The continuum model (\ref{energy}), however, becomes invalid for localized  
solutions with sizes of few lattice constants. In this region
we investigate solutions for chiral skyrmions within the discrete models. 
We consider classical spins, $\mathbf{S}_i$, of unit length on 
a two-dimensional square lattice with the following energy functional 
\cite{Leonov14a} $E = E_0 + E_D$ where
%
%
%
\begin {eqnarray}
E_0 = -J \,\sum_{<i,j>} (\mathbf{S}_i \cdot \mathbf{S}_j ) 
-\sum_{i} [\mathbf{H} \cdot \mathbf{S}_i 
+K ( \mathbf{S}_i \cdot \mathbf{n} )^2], \:
\label{discrete1}
\end{eqnarray}
and the Dzyaloshinskii-Moriya energy equals

\begin {eqnarray}
E_D = -D \, \sum_{i} \left(\mathbf{S}_i \times \mathbf{S}_{i+\hat{x}} \cdot \hat{x}
 + \mathbf{S}_i \times \mathbf{S}_{i+\hat{y}} \cdot \hat{y} \right) 
\label{discrete1b}
\end{eqnarray}
for Bloch-type modulations, and
\begin {eqnarray}
 E_D = -D \, \sum_{i}(\mathbf{S}_i \times \mathbf{S}_{i+\hat{x}} \cdot \hat{y} 
- \mathbf{S}_i \times \mathbf{S}_{i+\hat{y}} \cdot \hat{x})
\label{discrete1n}
\end{eqnarray}
for N\'eel-type modulations ($<i,j>$ denotes pairs of nearest-neighbor spins).

%
%

For a helix $ \mathbf{S}_i = (\cos \theta_i, \sin \theta_i, 0)$ propagating
along the $x$-axis at field and anisotropy ($\mathbf{H} = K =0$),  
model (\ref{discrete1}) is reduced to
\begin {eqnarray}
E = \sum_{i} \left[ -J \cos (\theta_i - \theta_{i+\hat{x}}) 
-D \sin (\theta_i - \theta_{i+\hat{x}}) \right],
\label{discrete2}
\end{eqnarray}
and yields the equilibrium period $ p_0 = 2 \pi/\arctan(D/J)$ ($p_0$ is the number of magnetic ions corresponding to $\Delta \theta = 2 \pi$).

\begin{figure}
\includegraphics[width=1.0\columnwidth]{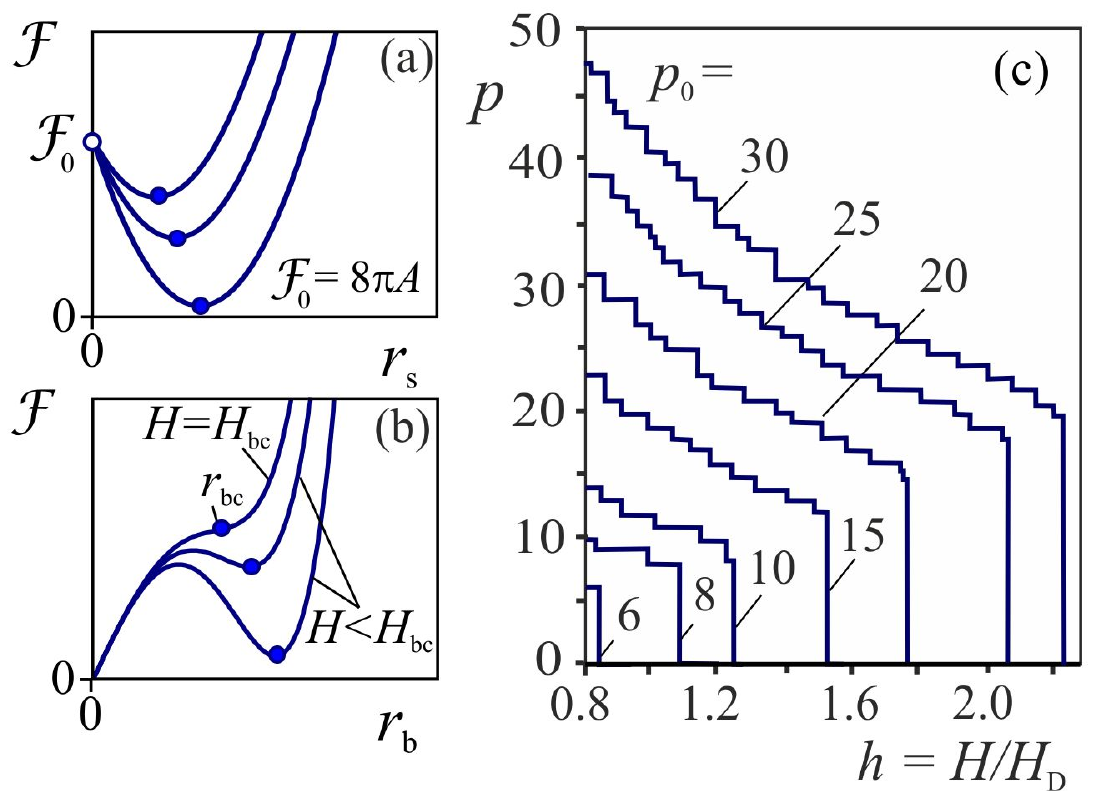}
\caption{
(color online). Micromagnetic energies of an isolated 
skyrmion (a) and a bubble domain (b) as function of 
their sizes for selected values of the applied magnetic 
fields \cite{Hubert98,JMMM94}. Isolated bubbles collapse
at critical field $H_{bc}$ with finite radius $r_{bc}$.
Isolated chiral skyrmions exist at very high fields
without collapse.
The equilibrium skyrmion sizes $p$ as functions
of the applied field calculated for different 
values of $p_0$ (\ref{discrete2})
indicate the collapse of chiral skyrmions (c) ($p$ is
defined here as a diameter of a circle encompassing
a skyrmion core area with $m_z \leq 0.995$).
\label{fig:discrete1}
}
\end{figure}

\begin{figure}
\includegraphics[width=1.0\columnwidth]{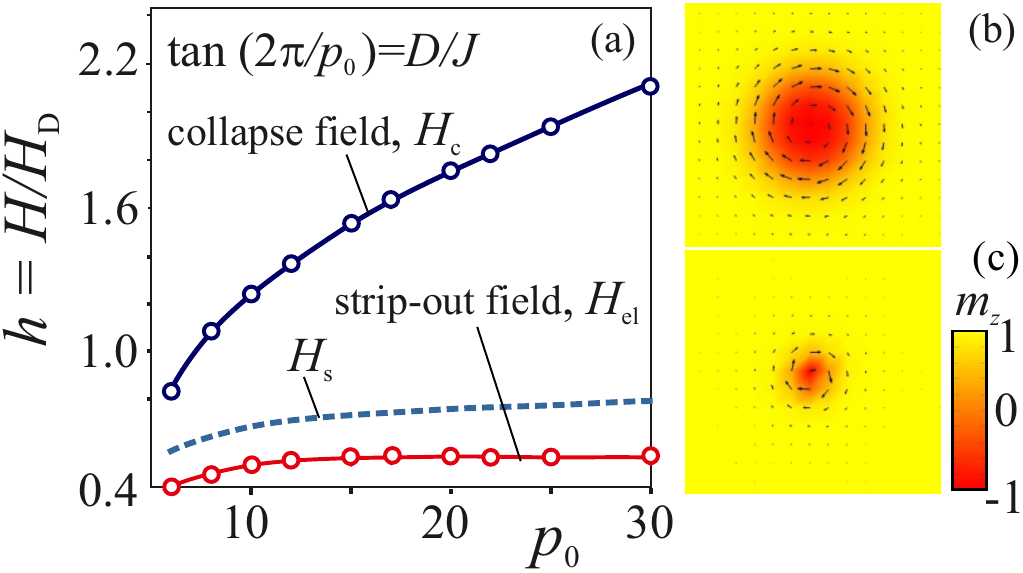}
\caption{
(color online). Collapse  ($H_c$) and strip-out or elliptic instability
 ($H_{el}$) critical fields calculated within  the discrete model 
(\ref{discrete1})  for $k = 0$ and different values of $p_0$ (\ref{discrete2}).
At the dashed line $H_s (p_0)$  the isolated skyrmion energy $\mathcal{F}$
equals zero, and below this line skyrmions can condense
into the hexagonal lattice (a).
Calculated distributions of the magnetization of the skyrmion core
for $k = 1.315$, $p_0 = 24$ and for $h = 0.42$ (b) and $h = 1.27$ (c).
\label{fig:discrete1a}
}
\end{figure}

The calculated equilibrium skyrmion diameter $p$ as a function of the applied
field indicates the collapse of the skyrmion
core at certain finite fields $h_c(p_0)$ (Fig. \ref{fig:discrete1} (a)).
The critical field $h_c(p_0)$  increases without limit with increasing $p_0$ 
(Fig. \ref{fig:discrete1} (b)) and, thus, signifies a transition from the discrete 
model to the continuous model.

\subsubsection{Elliptic instability (strip-out) at low fields }

Isolated skyrmions exist as metastable states above the critical field $h_s (k)$ 
(Fig. \ref{phasediagrams} (a), (b)). Below this line the energy
$\mathcal{F}$ (\ref{totalenergy}) becomes negative  and skyrmions tend to condense
into a hexagonal lattice \cite{JMMM94}. However, if the formation of skyrmion
lattices is suppressed (as in PdFe/Ir (111) films \cite{Romming15}) isolated skyrmions
continue to exist below the critical line $h_s (k)$ (with
the skyrmion core energy density lower than that of the surrounding 
saturated state). 
At the same time isolated skyrmions have a tendency to elongate 
and expand into a band with helicoidal or cycloidal modulations 
and eventually to fill the whole space, since the spiral state 
represents the minimum with lower energy as compared to the local 
minima with the metastable isolated skyrmions.
%
These (\textit{elliptic}) instabilities are similar to "strip-out" instabilities
of isolated magnetic bubbles at a certain critical field
\cite{Thiele70} observed in common ``bubble-domain'' films \cite{Hubert98}
and in magnetic nanolayers with perpendicular anisotropy \cite{Bran09}.
For chiral skyrmions,  the elliptic instability fields $H_{el}$ are calculated 
from the stability analysis of the skyrmion energy (\ref{totalenergy}) 
with respect to (elliptic) perturbations of type \cite{pss94}
\begin{eqnarray}
 \tilde{\rho} = \rho + \varepsilon \eta (\rho) \cos 2 \varphi, \quad
\tilde{\psi} = \psi + \zeta (\rho, \varphi), 
\label{ellipticmodes2}
\end{eqnarray}
($\varepsilon \ll 1$). For isolated Bloch-type skyrmions the calculated critical line $h_{el} (k)$
( $ 0 < k < k_a$ ) is plotted in Fig. \ref{phasediagrams}. 
These results are close to earlier calculations for stray-field free elliptical distortions 
(\ref{ellipticmodes2}) with ansatz functions $\eta (\rho) = \sin \theta /(1 + a \sin \theta)$ 
optimized with respect to the parameter $a$ \cite{pss94}. 

Within the discrete model (\ref{discrete1})
the critical field $h_{el}$ has been calculated for zero anisotropy ($k = 0$) and 
for $ 6 < p_0 < 30$.  Fig. \ref{fig:discrete1a} (a) shows that the strip-out field 
$h_{el}$ essentially decreases with the decreased size of skyrmions what can be 
beneficial for possible application of such skyrmions. However, the existence 
region of these isolated skyrmions is restricted by the lower field of collapse $h_c$. 

\subsubsection{The k - h phase diagrams}

In this section we consider the existence area for
isolated skyrmions in the magnetic phase diagram
(Fig. \ref{phasediagrams}).
The energy functional for uniaxial chiral ferromagnets
 (\ref{energy}) depends on the two independent 
control parameters, the reduced values of the applied field,
$h$ and uniaxial anisotropy, $k$ (\ref{units5}).
The magnetic phase diagram in variables $k$ and $h$ 
collects \textit{all} possible solutions for model
(\ref{energy}). The calculated phase diagram in  the inset
of Fig. \ref{phasediagrams} shows the existence areas 
of the cycloids and skyrmion lattices and the transition
lines between these modulated phases and the saturated
state.
The phase diagram indicates the critical fields at zero
anisotropy, the bicritical point $B$ (1.90, 0.10), and
 the critical point $A$ (2.67, 0) \cite{JMMM94} 
(for a detailed description of this phase diagram see
 Ref. \cite{Wilson14}).
Fig. \ref{phasediagrams} shows critical lines
for isolated skyrmions (results of the continuum model (\ref{energy})).
Isolated skyrmions condense into a skyrmion lattice 
when the applied magnetic field decreases to the critical value $h_s (k)$.
However, isolated skyrmions can exist as localized objects
below the critical line $h_s(k)$ and strip-out into helicoids
at the critical line $h_{el}$.

\begin{figure}
\includegraphics[width=1.0\columnwidth]{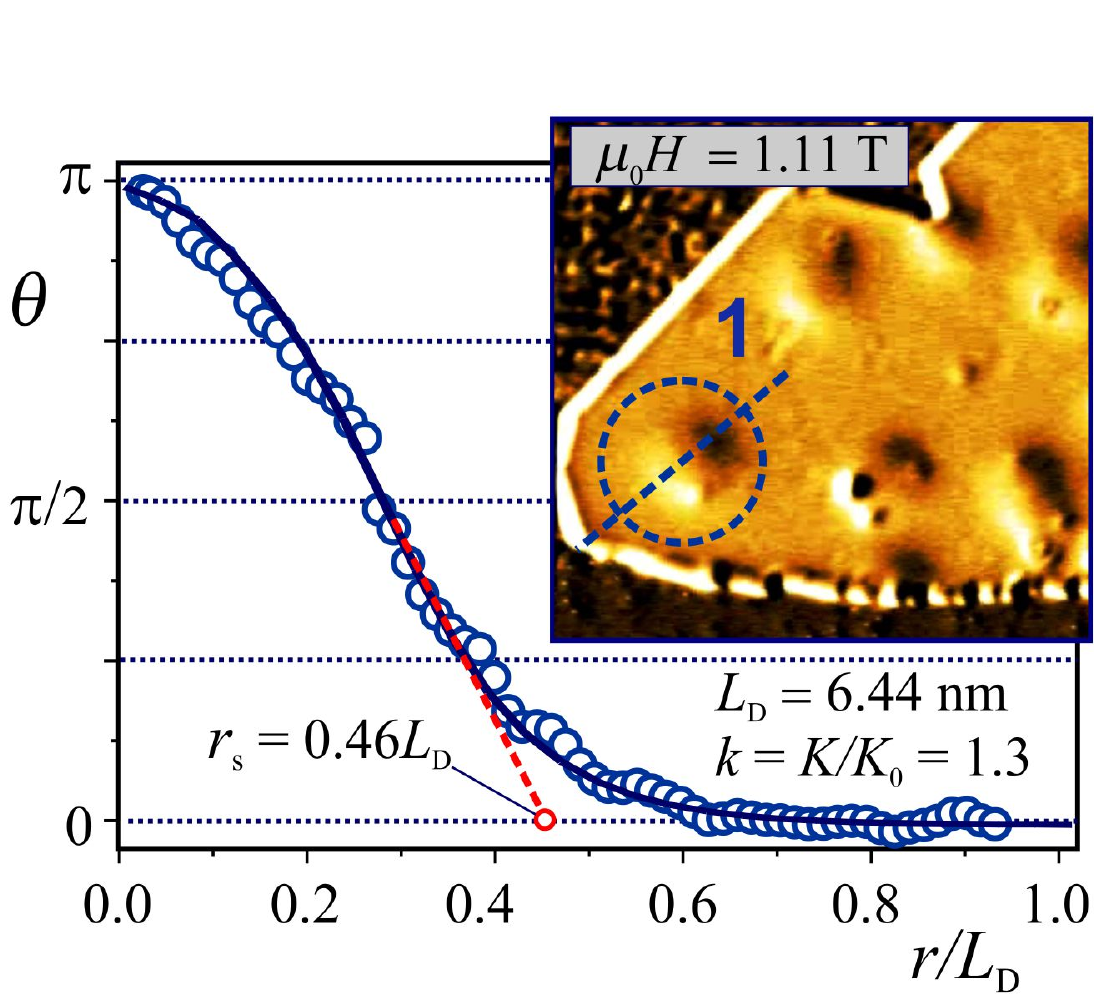}
\caption{
(color online). Magnetization profile $\theta (r)$ 
for an isolated skyrmion (1) (Inset) derived from  
SP-STM data for the applied field
$\mu_0 H = 1.11$ T.
The solid line is the solution of Eq. (\ref{equation1})
for $k = 1.315$ and $h = 0.321$, $r_s$ is the skyrmion core 
radius defined by Eq. (\ref{size}).
\label{fig:profiles2}
}
\end{figure}

\begin{figure*}
\includegraphics[width=2.0\columnwidth]{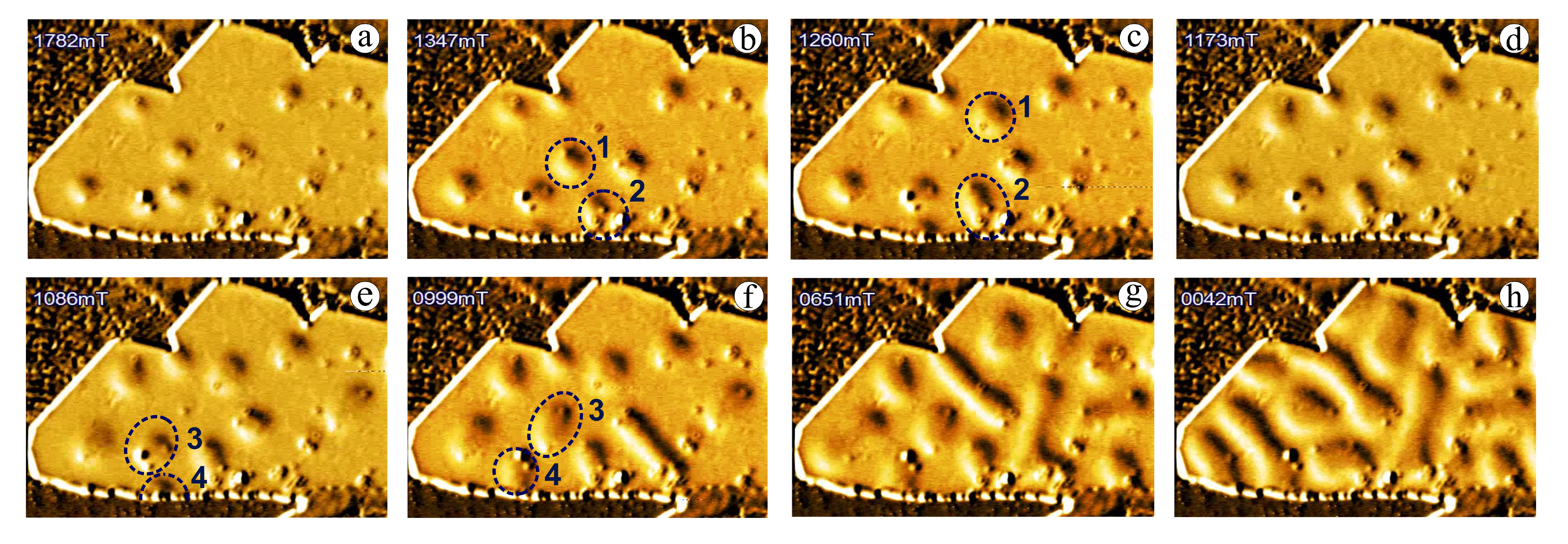}
\caption{
(color online). The extension of elliptic deformations in
isolated skyrmions in decreasing applied magnetic
fields (a-f) is terminated by the formation of
cycloid patterns (g,h).
\label{images1}
}
\end{figure*}

\section{ Experiment: isolated skyrmions in PdFe/Ir(111) nanolayers}

Sample preparation and spin-polarized (SP)- STM experiments were performed 
in a multi-chamber UHV system at a base pressure of 5e-11 mbar. 
Details of the sample preparation can be found in Ref. \cite{Romming13}. 
We use antiferromagnetic bulk Cr tips to minimize magnetostatic interactions 
between tip and sample. The SP-STM measurements were performed at T=4.2 
K in perpendicular magnetic fields of -3 to +3 Tesla. 
We repeatedly scanned the same sample area while continuously 
sweeping the magnetic field at a speed of ~12.8 mT/min, 
resulting in a series of images with a field difference of $\Delta B$=87 mT. 
Constant current images and maps of differential 
conductance ($dI/dU$) were measured simultaneously by a lock-in technique. 
We used small bias voltages ($U$ = 20 mV) and moderate currents 
($I$ = 3nA) to minimize the influence 
of the tunnel process on  the field-dependent magnetic evolution within 
the PdFe bilayer \cite{Romming13}.

PdFe/Ir(111) bilayers have a uniaxial anisotropy of  ``easy-axis'' type 
and exhibit chiral modulations of N$\acute{e}$el-type
\cite{Romming13,Romming15}.
It was also established by SP-STM observations that a cycloid  
(Fig. \ref{structure1a} (b)) is the ground
state of PdFe/Ir(111) films \cite{Romming13}.
 The material parameters of model (\ref{energy}) for PdFe/Ir(111) 
at $T = 4.2$ K determined in \cite{Dupe14,Romming15}
yield the following values for the characteristic parameters 
(\ref{unitsLD}): 
 $L_D = 6.44 = 23.85 a_0$ nm i.e. $p_0 = 24$ 
($a_0 =0.27$ nm is the lattice constant ), $\mu_0 H_D = 3.46$ T,
$K_0 = 1.9 \times 10^{6}$ J/m$^3$, 
$K_d = \mu_0 M^2/2 = 0.76  \times 10^{6}$ J/m$^3$ (``shape anisotropy'').
The sufficiently strong values of ``easy-axis''
anisotropy ($k = K/K_0 = 1.315$) ensures the stability 
of chiral modulations in PdFe/Ir(111) films  
with respect to stray-field effects \cite{Hubert98,Kiselev11} 
and make them convenient objects for investigations of chiral skyrmions
\cite{Romming13,Romming15}.

The calculated magnetic phase diagram of easy-axis chiral ferromagnets
includes the existence areas of one-dimensional modulations  and skyrmion 
lattices (Fig. \ref{phasediagrams}, Inset) \cite{JMMM94,Wilson14}.
These chiral modulations and transitions between
them  have been directly observed 
by Lorentz transmission electron microscopy (LTEM) 
in free standing  nanolayers  of cubic helimagnets 
\cite{Yu10,Yu12,Seki12,Tokunaga15} and
in PdFe/Ir(111) films by SP-STM \cite{Romming13}.

Isolated skyrmions and their internal structure
have been investigated by   SP-STM  in PdFe/Ir(111) bilayers
\cite{Romming13,Romming15}.
Following Ref. \cite{Romming15} we reconstruct the magnetization profile 
$\theta (r)$ for one of the isolated skyrmions in the film at the applied field 
$\mu_0 H = 1.11$ T (Fig. \ref{fig:profiles2}). These experimental 
results are in a close agreement with the solution of Eq. (\ref{equation1}) 
for $k =1.315$ and $h = 0.321$ (or $\mu_0 H = 1.11$ T.
In free-standing films of magnetically soft cubic helimagnets,
chiral skyrmions readily condense into hexagonal lattices
below $h_s(k)$ (Fig. \ref{phasediagrams}, inset) 
\cite{Yu10,Yu12,Seki12,Romming13}.
At low temperatures, however,  
an enhanced coercitivity of PdFe/Ir (111) bilayers
prevents the formation of skyrmion lattices below $h_s(k)$ 
(see the results of Ref. \cite{Romming15} for $T = 4.2$ K). 
This offers a unique opportunity to investigate 
isolated skyrmions in a broad range of
the applied fields.

\begin{figure}
\includegraphics[width=1.0\columnwidth]{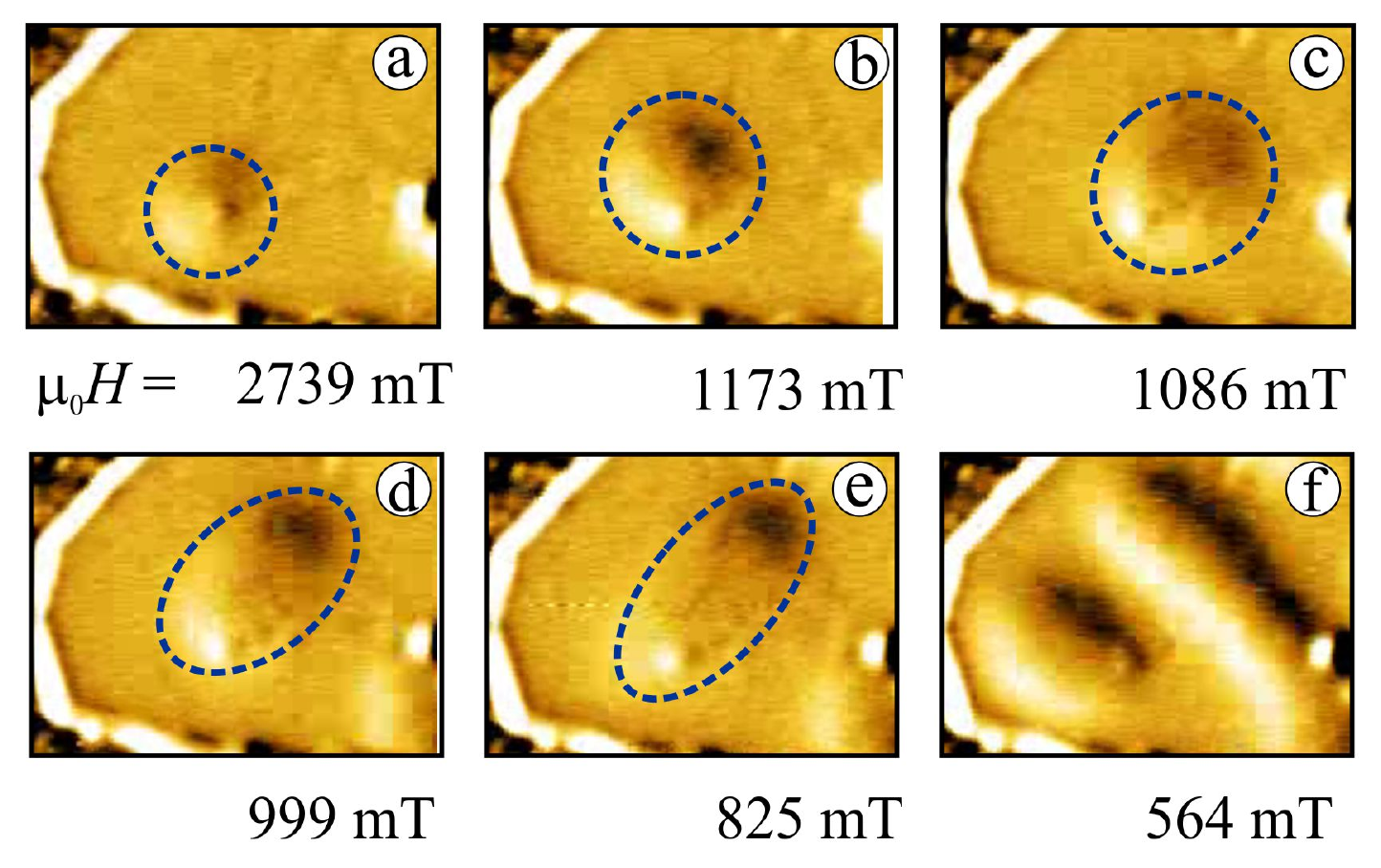}
\caption{
(color online). Isolated skyrmion core gradually increases
with decreasing field above the critical strip-out value $H_{el}$
(a,b) and stretches into a spiral domain for $H < H_{el}$ (c-f).
\label{fig:stripout}
}
\end{figure}

Figure \ref{images1} shows selected frames from the whole 
SP-STM data  set where the evolution from isolated skyrmions 
at high fields to spin spirals at zero fields can be observed. 
The two-lobe appearance of skyrmions is due to a predominantly 
in-plane magnetization of the Cr tip \cite{Romming13,Romming15}. 
The strip-out of skyrmions starts in Fig. \ref{images1} (c) 
where a skyrmion, labeled (1), has jumped to a different position 
and a skyrmion (2) has developed an elongated shape. 
In Fig. \ref{images1} (f) more skyrmions have adopted elongated 
shapes, a process that seems to be influenced 
and assisted by defects, see skyrmion (3), and the repulsive
interactions with other skyrmions and chiral modulations
along the sample edges (so called \textit{surface twists})
\cite{Meynell14,Rybakov13}. 
%
%
The strip-out process can be quantified more accurately 
in an area with only one strongly pinning defect, see detailed 
view in Fig. \ref{fig:stripout}. In Fig. \ref{images1} (c) the skyrmion 
shape starts to deviate from rotational symmetry  at $\mu_0 H ~1.10$ T
and becomes more and more elongated through Fig. \ref{images1} (d) and (e). 
Other skyrmions retain axial symmetry even at much lower fields.
The calculated value of the strip-out field for $k = 1.315$ equals
$\mu_0 H_{el} = 0.65$ T. The images in Figs. \ref{images1} 
(see also video materials in Ref. \cite{Romming15}) show that 
the elliptical instability field has different values
for different skyrmions and strongly  depends on  skyrmion-skyrmion
interactions, interactions with sample edges and defects.
Similar effects are characteristic for strip-out instabilities 
of isolated bubble domains (see e.g. \cite{Bran09}).

\begin{figure}
\includegraphics[width=0.9\columnwidth]{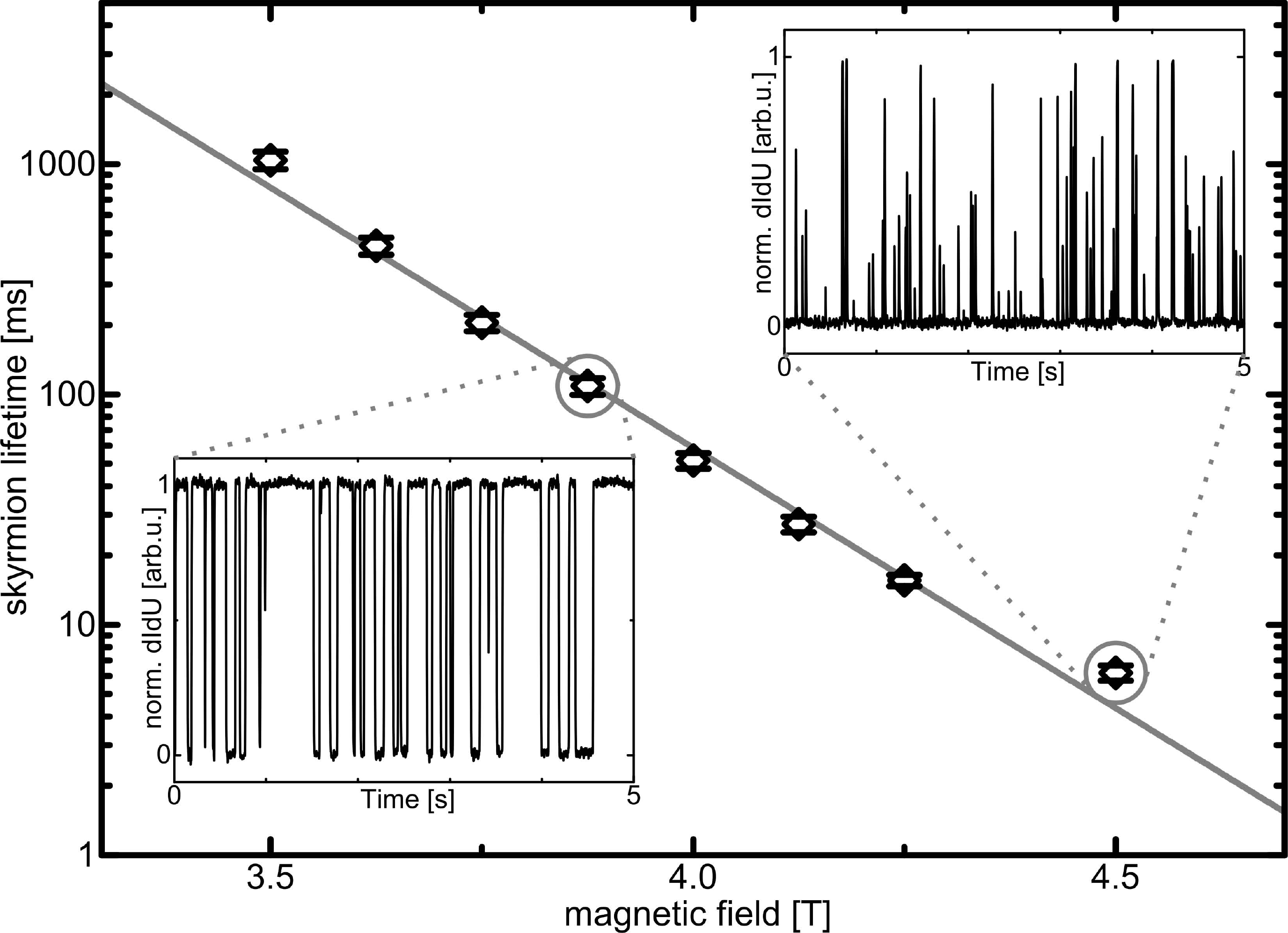}
\caption{Skyrmion lifetime as a function of external magnetic field. 
SP-STM tip is held stationary above a skyrmion position while tunneling 
with $U$ = +600 mV, $I$ = 100 nA. 
This induces a continuous, stochastic switching of the magnetic state under 
the tip between skyrmion state (1) and ferromagnetic state (0) (see resulting 
telegraph noise in insets). Data points show the average lifetime of the skyrmion 
state as derived from the telegraph noise signal. 
At 4.5 T, the lifetime approaches the time resolution limit of the STM while 
still detectable. Therefore the skyrmion state is still metastable and the 
collapse field must be higher than 4.5 T. ($T$ = 4.2 K)
\label{fig:collapse}
}
\end{figure}

To determine the skyrmion collapse field with reasonable 
statistical accuracy, we have monitored the repeated creation 
and annihilation of a skyrmion at higher tunnel bias and current 
as a function of applied field. With the tip positioned above 
the pinning defect, we monitored the telegraph noise in the 
spin-resolved $dI/dU$ signal and extracted the average lifetime 
of the skyrmion as a function 
of applied field (see insets in Fig. \ref{fig:collapse}).
The skyrmion lifetime decreases roughly exponentially 
\cite{Hagemeister15} down to a value of ~5 ms at 4.5 T.
Minimization of functional (\ref{discrete1}) with $p_0 = 24$ 
and $k = 1.315$ yields the collapse field 4.4 T
(cf. with collapse field of 7.1 Tesla 
calculated for zero anisotropy  (\ref{discrete1})).

\section{Conclusions}

Detailed SP-STM investigations of magnetic states in PdFe/Ir (111) thin films
and a comprehensive theoretical analysis within the standard  model (\ref{energy})
enable to describe the basic magnetic properties of isolated chiral skyrmions
and describe their evolution in a broad range of applied magnetic fields.

The equilibrium states of isolated axisymmetric skyrmions 
are described by differential equation (\ref{equation1}) \textit{common}
for different groups of chiral magnets \cite{JETP89,JMMM94}. 
Moreover, similar equations describe axisymmetric solitonic states
in other condensed matter systems with broken inversion symmetry 
\cite{Nature06,Leonov2014,Ackerman2014,Wright89}.
This implies a universal character  of chiral skyrmion properties and 
allows to consider the investigations in PdFe/Ir films as representative 
of the entire phenomenon. These investigations include general features 
of the chiral skyrmion evolution in the applied magnetic fields terminated 
at lower fields by the formation of skyrmion condensates or by elliptic 
instabilities of individual skyrmions and the collapse of the skyrmion 
core at high fields.

In this paper we have investigated magnetic properties of solitary skyrmions
only and neglected their interactions with other skyrmions, 
with chiral  modulations arising at the sample edges\cite{Meynell14},
and different types of defects.
We also have considered skyrmions magnetically homogeneous along their
axis. This assumption is justified by the structure of
PdFe/Ir (111) bilayers that exclude magnetic modulations along 
the film thickness \cite{JETP89}.
In thicker cubic helimagnets and uniaxial ferromagnets with $D_n$ 
and $C_n$ symmetry, however, such modulations are physically 
admissible and influence magnetic properties of these systems 
\cite{Rybakov13,Meynell14}.

\acknowledgements{
A.O.L acknowledges financial support by the FOM grant 11PR2928.
N.R., A.K., and R.W. acknowledge support  by the Deutsche 
Forschungsgemeinschaft via SFB668-A8, and A.N.B via Grant 
No. BO 4160/1-1. 
}

\appendix*

\section{}

\subsection{Skyrmion structure in different classes of uniaxial helimagnets}

Functional forms of $w_D$ energy contributions are determined by
crystallographic symmetry of a noncentrosymmetric magnetic crystal
\cite{Dz64,Bak80,JETP89}.

\begin{equation}
w_1^{(\pm)} = \mathcal{L}_{zx}^{(x)} \pm \mathcal{L}_{zy}^{(y)} ,
\quad
w_2^{(\pm)} = \mathcal{L}_{zx}^{(y)} \pm \mathcal{L}_{zy}^{(x)} ,
\label{lifshitz2}
\end{equation}

\begin{eqnarray}
&& C_{nv}:  [ w_1^{(+)}], \quad D_{2d}:  [ w_1^{(-)}], 
\quad   D_n: [ w_2^{(-)}, \mathcal{L}_{xy}^{(z)} ],\nonumber 
\\
&&  S_4: [w_1^{(-)}, w_2^{(-)}], \quad
C_n: [w_1^{(+)}, w_2^{(+)}, \mathcal{L}_{xy}^{(z)} ].
\label{lifshitz4}
\end{eqnarray}

For $\theta (\rho, \varphi)$, $\psi (\rho, \varphi)$

\begin{eqnarray}
&& w_1^{(+)} = \cos (\psi-\varphi) \theta_{\rho} 
- \sin \theta \cos \theta \sin (\psi- \varphi) \psi_{\rho}
\nonumber \\
&&-\frac{1}{\rho} \sin (\psi- \varphi) \theta_{\varphi}
+\frac{1}{\rho} \sin \theta \cos \theta \cos (\psi-\varphi) \psi_{\varphi}
\label{w1plus}
\end{eqnarray}
\begin{eqnarray}
&& w_1^{(-)} = \cos (\psi+\varphi) \theta_{\rho} 
- \sin \theta \cos \theta \sin (\psi+\varphi) \psi_{\rho}
\nonumber \\
&& -\frac{1}{\rho} \sin (\psi+ \varphi) \theta_{\varphi}
-\frac{1}{\rho} \sin \theta \cos \theta \cos (\psi+\varphi) \psi_{\varphi}
\label{w1minus}
\end{eqnarray}
\begin{eqnarray}
&& w_2^{(+)} = \sin (\psi-\varphi) \theta_{\rho} 
- \sin \theta \cos \theta \cos (\psi- \varphi) \psi_{\rho}
\nonumber \\
&&+\frac{1}{\rho} \cos (\psi- \varphi) \theta_{\varphi}
-\frac{1}{\rho} \sin \theta \cos \theta \sin (\psi-\varphi) \psi_{\varphi}
\label{w1plus1}
\end{eqnarray}
\begin{eqnarray}
&& w_2^{(-)} = \sin (\psi+\varphi) \theta_{\rho} 
- \sin \theta \cos \theta \cos (\psi+\varphi) \psi_{\rho}
\nonumber \\
&& +\frac{1}{\rho} \cos (\psi+ \varphi) \theta_{\varphi}
-\frac{1}{\rho} \sin \theta \cos \theta \cos (\psi+\varphi) \psi_{\varphi}
\label{w1minus1}
\end{eqnarray}

The solutions $\psi (\phi)$
are determined by crystal
classes of the system \cite{JETP89}.

\begin{eqnarray}
&& C_{nv}:   \psi = \varphi, \quad D_{2d}:  \psi = -\varphi + \pi/2,
\quad   D_n: \psi = \varphi + \pi/2,\nonumber 
\\
&&  S_4: \psi = -\varphi + \psi_1, \quad
C_n: \psi = \varphi + \psi_1.
\label{lifshitz3}
\end{eqnarray}
For ferromagnets belonging to $S_4$ and $C_n$ classes
energy functionals  $w_D$ include two terms:
$w_D = D_1 w_1^{(-)} + D_2 w_2^{(-)}$ for $S_4$
and
$w_D = D_1 w_1^{(+)} + D_2 w_2^{(+)}$ for $C_n$.
Angles $\psi_1 = \arctan (D_2/D_1)$ and the effective values of the 
DM constant are $D = \sqrt{D_1^2 + D_2^2}$.

For noncentrosymmetric cubic ferromagnets
belonging to  T and O  crystallographic classes
the energy functional $w_D$ has the following
form \cite{Bak80}
\begin{eqnarray}
w_D=\mathcal{L}_{yx}^{(z)}+\mathcal{L}_{xz}^{(y)}+\mathcal{L}_{zy}^{(x)}
=\mathbf{m}\cdot \mathrm{rot}\mathbf{m},
\label{Lifshitzc}
\end{eqnarray}
and stabilizes solutions with $\psi = \pi/2 + \varphi$.

The skyrmion energy densities for all these
structures can be reduced to 
a common functional form
\cite{JMMM94}.

\begin{figure*}
\includegraphics[width=2.0\columnwidth]{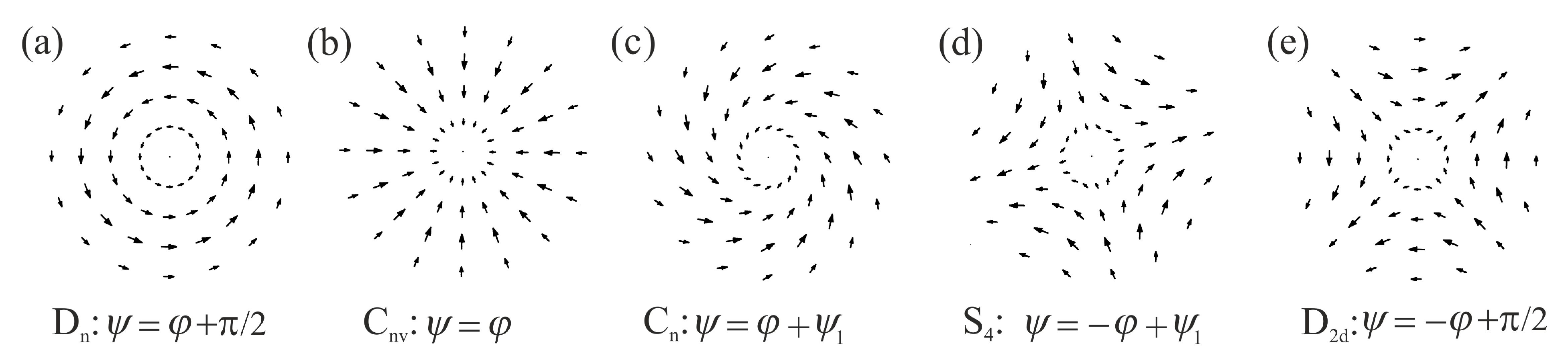}
\caption{
Projections of $\mathbf{m}$ onto the basal plane
in the core of chiral skyrmions for noncentrosymmetric uniaxial 
ferromagnets \cite{JETP89}.
\label{fig:structure2}
}
\end{figure*}

\subsection{Solutions for one-dimensional modulations}

For one-dimensional modulations propagating along the $\xi-$ axis
the energy functional (\ref{energy}) can be written as
\begin{eqnarray}
w\!=\!A(\theta_{\xi}^2\!+\!\sin^2\theta\psi_{\xi}^2)\!+\!w_D\!-
\!K\!\cos^2\!\theta\!-\!\mu_0MH\!\cos\!\theta. \qquad
\label{energyhelix}
\end{eqnarray}

In the DM energy contribution $w_D$ (\ref{energyhelix}) Lifshitz invariants  of type 
$\mathcal{L}_{ij}^{(x)}$, $\mathcal{L}_{ij}^{(y)}$ induce modulations propagating 
in the $xy$ plane
(e.g. \textit{helicoids} and \textit{cycloids} in Fig. 2 (a), (b)), and
invariants $\mathcal{L}_{xy}^{(z)}$ favour modulations along the $z$-axis
(\textit{cones}).

\textit{Helicoids and cycloids}. To be specific, we consider in-plane modulations
propagating along the $x-$ axis. Depending on the magnetic crystal symmetry, different
types of modulations are stabilized by the $w_D$ energy functional \cite{JETP89}.
Particularly, in cubic helimagnets and uniaxial ferromagnets of $D_n$  
crystallographic classes, $\mathbf{M}$ rotates as a Bloch-type domain 
wall (\textit{helicoids}), and in uniaxial ferromagnets with $C_{nv}$ 
symmetry the magnetization  rotates along the propagation direction 
like a N\'eel-type domain wall  (\textit{cycloids})
(Fig. \ref{structure1a} (a), (b)):

\begin{eqnarray}
&& \mathbf{m} \: = \: \vec{e}_y \: \sin  \theta \: (x) \ + \ \vec{e}_z \: \cos  \theta \: (x)
\quad \mathrm{(helicoids),}
\nonumber \\
&& \mathbf{m} \: = \: \vec{e}_y \: \sin  \theta \: (x) \ + \ \vec{e}_z \: \cos  \theta \: (x)
\quad \mathrm{(cycloids)}. \quad
\label{2helix}
\end{eqnarray}
The Euler equation for the functional 
\begin{eqnarray}
w_{\mathrm{h}} (\theta) = A \theta_x^2 - D \theta_x -\mu_0 MH \cos \theta
- K \cos^2 \theta \qquad
\label{energyhelix4}
\end{eqnarray}
yields  magnetization profiles  $\theta(x)$ for helicoids and cycloids
 \cite{Dz64}.
Analytical solutions for $\theta (x)$ describe helical modulations
distorted by the applied field and uniaxial anisotropy \cite{Dz64}.
These helicoids (cycloids) gradually unwind
into a set of isolated domain walls at critical line $H_{\mathrm{h}} (K)$ 
\cite{Dz64,JMMM94,Wilson14}.

\textit{Cones}.
In cubic helimagnets and uniaxial ferromagnets with $C_n$ and $D_n$ symmetries,
the DM energy functional $w_D$ includes Lifshitz invariants $\mathcal{L}_{xy}^{(z)}$
(\ref{lifshitz4}, \ref{Lifshitzc}) favouring chiral modulations (\textit{cone} phases)
along the $z-$ axis.
Minimization of the energy functional
\begin{eqnarray}
w_{\mathrm{h}} (\theta) = A \theta_z^2 - D \theta_z -\mu_0 MH \cos \theta 
- K \cos^2 \theta \qquad
\label{energycone}
\end{eqnarray}
yields the solutions for single-harmonic modulations describing the cone
phase \cite{Bak80,Butenko10}:
\begin{eqnarray}
\cos \theta = \frac{H}{H_D} \left( 1 - \frac{K}{K_0} \right)^{-1}, \quad \psi = z/L_D.
\label{cone1}
\end{eqnarray}
These equations include the characteristic parameters of a uniaxial chiral ferromagnet 
(\ref{unitsLD}).

\subsection{Characteristic lengths and critical parameters}

In uniaxial noncentrosymmetric ferromagnets, chiral modulations arise as a result of
a competition between the DM interactions favouring a rotation of the magnetization,
the exchange coupling and the ``potential'' energy $f(\theta) =-\mu_0 MH \cos \theta
- K \cos^2 \theta $ tending to suppress such modulations.
The balance between the chiral energy $w_D (\theta, \theta_{\rho})$ 
and the potential energy contributions $f(\theta)$ determine the equilibrium spin
configurations in chiral magnets. At zero field and for zero anisotropy 
single harmonic modulations $\theta = q_0 x$ ($q_0 = D/(2A)$) minimize the functional
(\ref{energyhelix}). In the opposite limit of strong anisotropy ($ K > \pi^2 D^2/(16 A)$)
these modulations transform into a set of isolated 180$^\circ$ domain walls separating
the homogeneous states with $\theta_1 = 0$, $\theta_2 = \pi$.

The width of an isolated Bloch domain wall 
$L_B$, its energy, $\gamma_B$, and anisotropy field $H_a$ are as follows:
\begin{eqnarray}
  L_B =  \pi \sqrt{\frac{A}{K}}, \quad \gamma_B = 4 \sqrt{A K}.
\quad  H_a = \frac{K}{\mu_0 M_0} \qquad
\label{units1}
\end{eqnarray}

These are the fundamental parameters describing magnetic states
in a common (centrosymmetric) uniaxial ferromagnet \cite{Hubert98}.
To demonstrate a competing character of the magnetic interactions
in chiral uniaxial ferromagnets, we consider an isolated domain wall 
at zero field that separates  the homogeneous states
with $\theta_1 =0$ and $\theta_2 = \pi$. The equilibrium
states of this domain wall are derived by minimization
of functional (\ref{energyhelix}) for $H =0$ \cite{Hubert98}.
The standard calculation of the wall energy \cite{Hubert98}  
$\gamma_{\mathrm{w}} = \int_0^{\infty} \left[ w_{\mathrm{h}}(\theta)
-w_{\mathrm{h}}(0) \right] dx$ 
yields the following result  \cite{JMMM94,PhysB05}

\begin{eqnarray}
 \gamma_{\mathrm{w}}  = 4 \sqrt{AK} - \pi |D| =
4 \pi A \left( \frac{1}{L_B} -\frac{\pi}{L_D} \right).
\label{wall1}
\end{eqnarray}
The first (positive) term in (\ref{wall1}) is the wall
energy of a uniaxial ferromagnet \cite{Hubert98}
arising as a common effect of the uniaxial anisotropy
pinning the magnetization vector along the easy-axis
and the exchange stiffness suppressing deviations
of $\mathbf{M}$ from these directions.
The negative energy contribution in $\gamma_{\mathrm{w}}$
(\ref{wall1}) is due to the DM interactions favouring
modulations of the magnetization with a specific rotation sense.
The strength of this ``winding force''
is characterized by $1/L_D$: the larger the DM
coupling, the smaller the period of the modulations.
For $L_D < \pi L_B$ the wall energy becomes negative
manifesting the instability of the homogeneous states
with respect to chiral modulations.

The dimensionless parameter
$\varkappa$ introduced as \cite{JMMM94,PhysB05}
\begin{eqnarray}
\gamma_{\mathrm{w}} = 
 4 \sqrt{AK} \left(1 - \varkappa \right), \quad 
\varkappa  = \frac{\pi}{4} \frac{|D|}{\sqrt{A K}}
 = \frac{\pi L_B}{L_D}, \qquad
\label{kappa} 
\end{eqnarray}
provides the criterion for the existence
of chiral modulated states.
For $ \varkappa >1$,  the DM interactions overcome
a pinning of the magnetization along easy-axis direction and stabilize
modulated states. For $ 0 < \varkappa <1$ chiral modulated phases are totally
suppressed, and chiral patterns exist as metastable localized states in a form
of isolated skyrmions and domain walls (kinks).

Parameter $\varkappa$ (\ref{kappa}) is similar to the 
\textit{Ginzburg-Landau parameter} 
$\varkappa_{GL} = \lambda/\xi$ in the theory of
Abrikosov vortices (the mixed state) in superconductors
 \cite{Abrikosov57}.
The parameter $\varkappa_{GL}$ is a ratio of two characteristic lengths,
the \textit{coherence length} $\xi$ and the \textit{penetration depth}
$\lambda$. Abrikosov vortices exist in superconductors with
 $\varkappa_{GL} > 1/\sqrt{2}$ (\textit{type-II superconductors}) 
\cite{Abrikosov57,DeGennes}.
Physical analogies between  superconductor's mixed states and
chiral magnetic modulations are discussed in \cite{JMMM94,PhysB05}.
The characteristic lengths $L_D$ and $L_B$ provide different ways to
introduce reduced variables into model (\ref{energy3}).

The reduced energy functional based on the characteristic length $L_B$ 
with control parameters $\tilde{h} = H/H_a$ and $\varkappa$ (\ref{kappa}) 
and the reduced magnetic field  (where $H_a$ define in Eq. (\ref{units1}) 
is  convenient for investigations of isolated skyrmions in helimagnets 
with a strong uniaxial anisotropy 
($0 < \varkappa < 1$) \cite{pss94,Kiselev11}.


\begin{thebibliography} {99}
 
\bibitem{Dz64}  
I.\ E.\ Dzyaloshinskii, Sov.\ Phys.\ JETP {\textbf{19}}, 960 (1964); {\textbf{20}}, 665 (1964).

\bibitem{JETP89}
A. N. Bogdanov and D. A. Yablonskii, Sov. Phys. JETP \ {\textbf 68}, 101 (1989).
 
\bibitem{JMMM94}
A. Bogdanov and A. Hubert, J. Magn. Magn. Mater. {\textbf{138}}, 255 (1994); {\textbf{195}}, 182 (1999).


\bibitem{JPCS11}
U. K. R\"{o}{\ss}ler, A. A. Leonov, and A. N. Bogdanov, J. Phys.: Conf. Ser. {\textbf{303}}, 012105 (2011).

\bibitem{Rajaraman87}
R. Rajaraman, \textit{Solitons and Instantons} (North Holland, Amsterdam, 1987).

\bibitem{Manton04}
N. Manton and P. Sutcliffe, \textit{Topological Solitons} (Cambridge University Press, 2004).

\bibitem{Skyrmion10}
 G. E. Brown and M. Rho (editors), \textit{The Multifaceted Skyrmion} (World Scientific, 2010).

\bibitem{Melcher15}
C. Melcher, Proc. R. Soc. A, {\textbf{470}}, 20140394  (2014).

\bibitem{Derrick64} G. H. Derrick, J. Math. Phys. {\textbf{5}}, 1252 (1964).



\bibitem{Skyrme62}
T. H. Skyrme, Proc. R. Soc. A, {\textbf{260}}, 127  (1961).

\bibitem{Leonov15a}
A. O. Leonov, M. Mostovoy,  Nat. Commun.  in press  (2015).


\bibitem{JETPL95} A. Bogdanov,   JETP Lett.  {\textbf 62}, 247 (1995).

\bibitem{Nature06} U. K. R\"{o}{\ss}ler, A. N. Bogdanov, C. Pfleiderer, 
Nature (London), {\textbf{442}}, 797 (2006).

\bibitem{Wilhelm11} H. Wilhelm, M. Baenitz, M. Schmidt, U. K. Roessler, A. A. Leonov, A. N. Bogdanov,  Phys. Rev. Lett. \textbf{107}, 127203 (2011). 

\bibitem{Wilson14}  M. N. Wilson, A. B. Butenko, A. N. Bogdanov, and T. L. Monchesky,  
Phys. Rev. B \textbf{89}, 094411 (2014).

\bibitem{Wilhelm12} H. Wilhelm, M. Baenitz, M. Schmidt, C. Naylor, R. Lortz, U. K. Roessler, A. A. Leonov,  and A. N. Bogdanov, J. Phys.: Condens. Matter \textbf{24}, 294204 (2012). 

\bibitem{Keesman15} R. Keesman, A. O. Leonov, S. Buhrandt, G. T. Barkema, L. Fritz, R. A. Duine, arXiv: 1506.00271.

\bibitem{Lamago06}
D. Lamago, R. Georgii, C. Pfleiderer, and P. B\"oni, 
Physica B {\textbf{385-386}}, 385 (2006).

\bibitem{Pappas09} C. Pappas,  E. Lelievre-Berna, P. Falus, P. M. Bentley, E. Moskvin, S. Grigoriev, P. Fouquet, and B. Farago,  Phys. Rev. Lett.  {\textbf{102}}, 197202 (2009).

\bibitem{Muhlbauer09} S. M\"uhlbauer, B. Binz, F. Jonietz, C. Pfleiderer, A. Rosch, A. Neubauer, R. Georgii, and P. B\"oni, Science, {\textbf{323}}, 915 (2009).

\bibitem{Yu10}
X. Z. Yu, Y. Onose, N. Kanazawa, J. H. Park, J. H. Han, Y. Matsui, N. Nagaosa, and Y. Tokura,  Nature (London), {\textbf{465}}, 901 (2010).

\bibitem{Yu12} 
X. Z. Yu,  N. Kanazawa, Y. Onose, K. Kimoto, W. Z. Zhang, S. Ishiwata, Y. Matsui, and Y. Tokura,  Nat. Mater. \textbf{10}, 106 (2011); 

X. Z. Yu, N.  Kanazawa, W. Z.  Zhang, T.  Nagai, T.  Hara, K.  Kimoto, Y. Matsui, Y. Onose, Y. Tokura,  Nat. Commun. \textbf{3}, 988 (2012); 
A. Tonomura, X. Yu, K. Yanagisawa, T. Matsuda, Y. Onose, N. Kanazawa, H. S. Park, and Y. Tokura,  Nano Lett. \textbf{12}, 1673 (2012).

\bibitem{Wilson12} M. N. Wilson, E. A. Karhu, A. S. Quigley, U. K. R\"{o}{\ss}ler, A. B. Butenko, A. N. Bogdanov, M. D. Robertson, and T. L. Monchesky,  Phys. Rev. B \textbf{86}, 144420 (2012).

\bibitem{Huang12} S. X. Huang and  C. L. Chien,  Phys. Rev. Lett. {\textbf{108}}, 267201 (2012).

\bibitem{Heinze11} S. Heinze, K. von Bergmann, M. Menzel, J. Brede, A. Kubetzka, R.  Wiesendanger,  G. Bihlmayer, S. Bl\"ugel, Nat. Phys. {\textbf{7}}, 713 (2011).
	
\bibitem{Kezsmarki15}  I. K\'ezsm\'arki, S. Bord\'acs, P. Milde, E. Neuber, L. M. Eng, J. S. White, H. M. Rønnow, C. D. Dewhurst, M. Mochizuki, K. Yanai, H. Nakamura, D. Ehlers, V. Tsurkan, A. Loidl,  arXiv:1502.08049.

\bibitem{Tokunaga15} Y. Tokunaga, X. Z. Zu, J. S. White, H. M. Ronnow, D. Morikawa, Y. Taguchi, and Y. Tokura,  Nat. Commun.  {\textbf{6}}, 7638 (2015).

\bibitem{Seki12}
S. Seki, X. Z. Yu, S. Ishiwata, and Y. Tokura, 
Science \textbf{336}, 198 (2012).


\bibitem{Romming13} N. Romming, C. Hanneken, M. Menzel, J. E. Bickel, B. Wolter, K. von Bergmann, A. Kubetzka, and R. Wiesendanger,   Science {\textbf{341}}, 636 (2013).
	
\bibitem{Romming15}  N. Romming, A. Kubetzka, C. Hanneken, K. v. Bergmann, R. Wiesendanger,  Phys. Rev. Lett.  {\textbf{114}}, 177203 (2015); 	C. H. Marrows,  Physics, {\textbf{8}}, 40 (2015).

\bibitem{pss94}
A. Bogdanov and A. Hubert,  Phys. Stat. Sol. (b)
{\textbf{138}}, 255 (1994), {\textbf{186}}, 527 (1994).

\bibitem{Kiselev11}
N. S. Kiselev, A. N. Bogdanov, R. Sch\"{a}fer, and U.K.R\"{o}{\ss}ler,  J. Phys. D: Appl. Phys. {\textbf{44}}, 392001 (2011).

\bibitem{Hubert98}
A. Hubert, R. Sch\"{a}fer, \textit{Magnetic Domains} (Springer, Berlin, 1998).

\bibitem{Tu71} Y. Tu,  J. Appl. Phys. {\textbf{42}}, 5704  (1971).

\bibitem{Bergmann15} K. von Bergmann, M. Menzel, A. Kubetzka, and R. Wiesendanger,   Nano Lett. {\textbf{15}}, 3280 (2015).

\bibitem{Butenko10} A. B. Butenko, A. A. Leonov, U. K. R\"{o}{\ss}ler, A. N. Bogdanov,  Phys. Rev. B {\textbf{82}}, 052403 (2010).

\bibitem{Lin13} S. Z. Lin,C.Reichhardt, C. D. Batista, and A. Saxena,  
Phys. Rev. B {\textbf{87}}, 214419 (2013).

\bibitem{Rohart13} S. Rohart and A. Thiaville,  Phys. Rev. B {\textbf{88}}, 184422 (2013).

\bibitem{Kim14} J.-V. Kim, F. Garcia-Sanchez, J. Sampaio, C. Moreau-Luchaire, V. Cros, and A. Fert,  Phys. Rev. B {\textbf{90}}, 064410 (2014).

\bibitem{Sampaio13}
J. Sampaio, V. Cross, S. Rohart, A. Thiaville, and A. Fert,   Nat. Nanotech. {\textbf{8}}, 839 (2013).

\bibitem{Leonov14a}
A. O. Leonov, U. K. R\"{o}{\ss}ler, M. Mostovoy, 
EPJ: Web of Conf.  {\textbf{75}}, 05002 (2014).

\bibitem{Butenko09}
A. B. Butenko, A. A. Leonov, A. N. Bogdanov, and U. K. R\"{o}{\ss}ler,  Phys. Rev. B {\textbf{80}}, 134410 (2009);

\bibitem{Iwasaki13}
J. Iwasaki, M. Mochizuki, N. Nagaosa, 
Nat. Nanotech. {\textbf{8}}, 742 (2013).

\bibitem{Zhang15}
X. Zhang, G. P. Zhao, H. Fangohr, J. P. Liu, W. X. Xia, F. J. Morvan,  Sc. Reports {\textbf{5}}, 7643 (2015).

\bibitem{Nagaosa13}
N. Nagaosa and Y. Tokura, Nat. Nanotech. {\textbf{8}}, 899 (2013).


\bibitem{Belavin75} A. A. Belavin and A. M. Polyakov,  Sov. Phys. JETP Lett. \ {\textbf 22}, 245 (1975).

\bibitem{Thiele70} A. A. Thiele,  J. Appl. Phys. {\textbf{41}}, 1139  (1970).

\bibitem{Bran09} C. Bran, A. B.  Butenko, N. S.  Kiselev, U.  Wolff, L. Schultz, O.  Hellwig, U. K. Roessler, A. N. Bogdanov, V. Neu,  Phys. Rev. B,  
\textbf{79}, 024430 (2009).

\bibitem{Dupe14}
B. Dupe, M.Hoffmann, C. Paillard, and S. Heinze, 
Nat. Commun.  \textbf{5}, 4030 (2014).


\bibitem{Meynell14}
S. A. Meynell, M. N. Wilson, H. Fritzsche, A. N. Bogdanov, and T. L. Monchesky, 
Phys. Rev. B {\textbf{90}}, 014406 (2014);

M. N. Wilson, E. A. Karhu, D. P. Lake, A. S. Quigley, A. N. Bogdanov, U. K. Rößler, T. L. Monchesky,  Phys. Rev. B {\textbf{88}}, 214420 (2013).

\bibitem{Rybakov13}
F. N. Rybakov, A. B. Borisov, A. N. Bogdanov,  Phys. Rev. B {\textbf{87}}, 094424 (2013).

\bibitem{Hagemeister15}
J. Hagemeister et al. Nat. Commun.  in press (2015).

\bibitem{Leonov2014}
A. O. Leonov, I. E. Dragunov, U. K. R\"{o}{\ss}ler , and A. N. Bogdanov, 
Phys. Rev. E,  \textbf{90}, 042504 (2014).

\bibitem{Ackerman2014} P. J. Ackerman, R. P. Trivedi, B. Senyuk, J. van de Lagemaat, and I. I. Smalyukh,  Phys. Rev. E,  \textbf{90}, 012505 (2014).


\bibitem{Wright89} G. H. Wright and N. D. Mermin, Rev. Mod. Phys. {\textbf{61}}, 385 (1989).

\bibitem{Bak80}
P.\ Bak and M.\ H.\ Jensen, J.\ Phys. C: Solid State Phys.
\ {\textbf 13}, L881 (1980).

\bibitem{PhysB05}
A. N. Bogdanov,  U. K. R\"o\ss ler, C. Pfleiderer,  Physica B \textbf{359-61}, 1162 (2005).

\bibitem{Abrikosov57}
A. A. Abrikosov, Sov. Phys. JETP {\textbf{5}}, 1174 (1957).

\bibitem{DeGennes}
P. G. de Gennes, \textit{Superconductivity of Metals and Alloys } (W. A. Benjamin, New York, 1966).















\end{thebibliography}
\end{document}